

\def\singlespace{\baselineskip=\normalbaselineskip}

\def\oneandahalfspace{\baselineskip=\normalbaselineskip
  \multiply\baselineskip by 3 \divide\baselineskip by 2}

\parskip=\medskipamount
\overfullrule=0pt
\raggedbottom
\def\normalparindent{24pt}
\nopagenumbers
\footline={\ifnum\pageno=1{\hfil}\else{\hfil\rm\folio\hfil}\fi}
\def\endpage{\vfill\eject}
\def\beginlinemode{\endmode\begingroup\parskip=0pt
                   \obeylines\def\\{\par}\def\endmode{\par\endgroup}}
\def\beginparmode{\endmode\begingroup \def\endmode{\par\endgroup}}
\let\endmode=\par
\def\raggedcenter{
                  \leftskip=2em plus 6em \rightskip=\leftskip
                  \parindent=0pt \parfillskip=0pt \spaceskip=.3333em
                  \xspaceskip=.5em\pretolerance=9999 \tolerance=9999
                  \hyphenpenalty=9999 \exhyphenpenalty=9999 }
\def\\{\cr}
\let\rawfootnote=\footnote\def\footnote#1#2{{\parindent=0pt\parskip=0pt
        \rawfootnote{#1}{#2\hfill\vrule height 0pt depth 6pt width 0pt}}}
\def\title{\null\vskip 3pt plus 0.2fill\beginlinemode\raggedcenter\bf}
\def\author{\vskip 3pt plus 0.2fill \beginlinemode\raggedcenter}
\def\affil{\vskip 3pt plus 0.1fill\beginlinemode\raggedcenter\it}
\def\abstract{\vskip 3pt plus 0.3fill \beginparmode{
\centerline {ABSTRACT}\medskip \noindent ~}  }
\def\endtitlepage{\endpage\body}
\def\body{\beginparmode\parindent=\normalparindent}
\def\head#1{\par\goodbreak{\immediate\write16{#1}
           {\noindent\bf #1}\par}\nobreak\nobreak}

\def\refto#1{$^{-#1-}$}
\def\ref#1{Ref.~#1}
\def\cite#1{{#1}}\def\/#1/{[\cite{#1}]}
\def\Ref#1{Ref.$\ [#1]$}

\def\(#1){(\call{#1})}
\def\call#1{{#1}}\def\taghead#1{{#1}}
\def\references{\head{REFERENCES}\beginparmode\frenchspacing\parskip=0pt}
\gdef\refis#1{\item{#1.\ }}
\gdef\journal#1,#2,#3,#4.{{\sl #1~}{\bf #2}, #3 (#4)}
\def\endreferences{\body}
\def\endit{\endmode\vfill\supereject}\let\endpaper=\endit



\def\gsim{\mathrel{\raise.3ex\hbox{$>$\kern-.75em\lower1ex\hbox{$\sim$}}}}
\def\lsim{\mathrel{\raise.3ex\hbox{$<$\kern-.75em\lower1ex\hbox{$\sim$}}}}
\def\square{\kern1pt\vbox{\hrule height 0.6pt\hbox{\vrule width 0.6pt\hskip 3pt
   \vbox{\vskip 6pt}\hskip 3pt\vrule width 0.6pt}\hrule height 0.6pt}\kern1pt}
\def\sla{\raise.15ex\hbox{$/$}\kern-.72em}

\catcode`@=11
\newcount\r@fcount \r@fcount=0\newcount\r@fcurr
\immediate\newwrite\reffile\newif\ifr@ffile\r@ffilefalse
\def\w@rnwrite#1{\ifr@ffile\immediate\write\reffile{#1}\fi\message{#1}}
\def\writer@f#1>>{}
\def\referencefile{\r@ffiletrue\immediate\openout\reffile=\jobname.ref%
  \def\writer@f##1>>{\ifr@ffile\immediate\write\reffile%
    {\noexpand\refis{##1} = \csname r@fnum##1\endcsname = %
     \expandafter\expandafter\expandafter\strip@t\expandafter%
     \meaning\csname r@ftext\csname r@fnum##1\endcsname\endcsname}\fi}%
  \def\strip@t##1>>{}}

\def\citeall#1{\xdef#1##1{#1{\noexpand\cite{##1}}}}
\def\cite#1{\each@rg\citer@nge{#1}}
\def\each@rg#1#2{{\let\thecsname=#1\expandafter\first@rg#2,\end,}}
\def\first@rg#1,{\thecsname{#1}\apply@rg}
\def\apply@rg#1,{\ifx\end#1\let\next=\relax%
\else,\thecsname{#1}\let\next=\apply@rg\fi\next}%
\def\citer@nge#1{\citedor@nge#1-\end-}
\def\citer@ngeat#1\end-{#1}
\def\citedor@nge#1-#2-{\ifx\end#2\r@featspace#1
  \else\citel@@p{#1}{#2}\citer@ngeat\fi}
\def\citel@@p#1#2{\ifnum#1>#2{\errmessage{Reference range #1-#2\space is bad.}
    \errhelp{If you cite a series of references by the notation M-N, then M and
    N must be integers, and N must be greater than or equal to M.}}\else%
{\count0=#1\count1=#2\advance\count1 by1\relax\expandafter\r@fcite\the\count0,%
  \loop\advance\count0 by1\relax
    \ifnum\count0<\count1,\expandafter\r@fcite\the\count0,%
  \repeat}\fi}
\def\r@featspace#1#2 {\r@fcite#1#2,}    \def\r@fcite#1,{\ifuncit@d{#1}
    \expandafter\gdef\csname r@ftext\number\r@fcount\endcsname%
    {\message{Reference #1 to be supplied.}\writer@f#1>>#1 to be supplied.\par
     }\fi\csname r@fnum#1\endcsname}
\def\ifuncit@d#1{\expandafter\ifx\csname r@fnum#1\endcsname\relax%
\global\advance\r@fcount by1%
\expandafter\xdef\csname r@fnum#1\endcsname{\number\r@fcount}}
\let\r@fis=\refis   \def\refis#1#2#3\par{\ifuncit@d{#1}%
    \w@rnwrite{Reference #1=\number\r@fcount\space is not cited up to now.}\fi%
  \expandafter\gdef\csname r@ftext\csname r@fnum#1\endcsname\endcsname%
  {\writer@f#1>>#2#3\par}}
\def\r@ferr{\endreferences\errmessage{I was expecting to see
\noexpand\endreferences before now;  I have inserted it here.}}
\let\r@ferences=\references
\def\references{\r@ferences\def\endmode{\r@ferr\par\endgroup}}
\let\endr@ferences=\endreferences
\def\endreferences{\r@fcurr=0{\loop\ifnum\r@fcurr<\r@fcount
    \advance\r@fcurr by 1\relax\expandafter\r@fis\expandafter{\number\r@fcurr}%
    \csname r@ftext\number\r@fcurr\endcsname%
  \repeat}\gdef\r@ferr{}\endr@ferences}
\let\r@fend=\endpaper\gdef\endpaper{\ifr@ffile
\immediate\write16{Cross References written on []\jobname.REF.}\fi\r@fend}
\catcode`@=12
\citeall\refto\citeall\ref\citeall\Ref
\catcode`@=11
\newcount\tagnumber\tagnumber=0
\immediate\newwrite\eqnfile\newif\if@qnfile\@qnfilefalse
\def\write@qn#1{}\def\writenew@qn#1{}
\def\w@rnwrite#1{\write@qn{#1}\message{#1}}
\def\@rrwrite#1{\write@qn{#1}\errmessage{#1}}
\def\taghead#1{\gdef\t@ghead{#1}\global\tagnumber=0}
\def\t@ghead{}\expandafter\def\csname @qnnum-3\endcsname
  {{\t@ghead\advance\tagnumber by -3\relax\number\tagnumber}}
\expandafter\def\csname @qnnum-2\endcsname
  {{\t@ghead\advance\tagnumber by -2\relax\number\tagnumber}}
\expandafter\def\csname @qnnum-1\endcsname
  {{\t@ghead\advance\tagnumber by -1\relax\number\tagnumber}}
\expandafter\def\csname @qnnum0\endcsname
  {\t@ghead\number\tagnumber}
\expandafter\def\csname @qnnum+1\endcsname
  {{\t@ghead\advance\tagnumber by 1\relax\number\tagnumber}}
\expandafter\def\csname @qnnum+2\endcsname
  {{\t@ghead\advance\tagnumber by 2\relax\number\tagnumber}}
\expandafter\def\csname @qnnum+3\endcsname
  {{\t@ghead\advance\tagnumber by 3\relax\number\tagnumber}}
\def\equationfile{\@qnfiletrue\immediate\openout\eqnfile=\jobname.eqn%
  \def\write@qn##1{\if@qnfile\immediate\write\eqnfile{##1}\fi}
  \def\writenew@qn##1{\if@qnfile\immediate\write\eqnfile
    {\noexpand\tag{##1} = (\t@ghead\number\tagnumber)}\fi}}
\def\callall#1{\xdef#1##1{#1{\noexpand\call{##1}}}}
\def\call#1{\each@rg\callr@nge{#1}}
\def\each@rg#1#2{{\let\thecsname=#1\expandafter\first@rg#2,\end,}}
\def\first@rg#1,{\thecsname{#1}\apply@rg}
\def\apply@rg#1,{\ifx\end#1\let\next=\relax%
\else,\thecsname{#1}\let\next=\apply@rg\fi\next}
\def\callr@nge#1{\calldor@nge#1-\end-}\def\callr@ngeat#1\end-{#1}
\def\calldor@nge#1-#2-{\ifx\end#2\@qneatspace#1 %
  \else\calll@@p{#1}{#2}\callr@ngeat\fi}
\def\calll@@p#1#2{\ifnum#1>#2{\@rrwrite{Equation range #1-#2\space is bad.}
\errhelp{If you call a series of equations by the notation M-N, then M and
N must be integers, and N must be greater than or equal to M.}}\else%
{\count0=#1\count1=#2\advance\count1 by1\relax\expandafter\@qncall\the\count0,%
  \loop\advance\count0 by1\relax%
    \ifnum\count0<\count1,\expandafter\@qncall\the\count0,  \repeat}\fi}
\def\@qneatspace#1#2 {\@qncall#1#2,}
\def\@qncall#1,{\ifunc@lled{#1}{\def\next{#1}\ifx\next\empty\else
  \w@rnwrite{Equation number \noexpand\(>>#1<<) has not been defined yet.}
  >>#1<<\fi}\else\csname @qnnum#1\endcsname\fi}
\let\eqnono=\eqno\def\eqno(#1){\tag#1}\def\tag#1$${\eqnono(\displayt@g#1 )$$}
\def\aligntag#1\endaligntag  $${\gdef\tag##1\\{&(##1 )\cr}\eqalignno{#1\\}$$
  \gdef\tag##1$${\eqnono(\displayt@g##1 )$$}}
\def\eqalignno#1{\displ@y \tabskip\centering
  \halign to\displaywidth{\hfil$\displaystyle{##}$\tabskip\z@skip
    &$\displaystyle{{}##}$\hfil\tabskip\centering
    &\llap{$\displayt@gpar##$}\tabskip\z@skip\crcr
    #1\crcr}}
\def\displayt@gpar(#1){(\displayt@g#1 )}
\def\displayt@g#1 {\rm\ifunc@lled{#1}\global\advance\tagnumber by1
        {\def\next{#1}\ifx\next\empty\else\expandafter
        \xdef\csname @qnnum#1\endcsname{\t@ghead\number\tagnumber}\fi}%
  \writenew@qn{#1}\t@ghead\number\tagnumber\else
        {\edef\next{\t@ghead\number\tagnumber}%
        \expandafter\ifx\csname @qnnum#1\endcsname\next\else
        \w@rnwrite{Equation \noexpand\tag{#1} is a duplicate number.}\fi}%
  \csname @qnnum#1\endcsname\fi}
\def\ifunc@lled#1{\expandafter\ifx\csname @qnnum#1\endcsname\relax}
\let\@qnend=\end\gdef\end{\if@qnfile
\immediate\write16{Equation numbers written on []\jobname.EQN.}\fi\@qnend}
\catcode`@=12


%
\newbox\hdbox%
\newcount\hdrows%
\newcount\multispancount%
\newcount\ncase%
\newcount\ncols
\newcount\nrows%
\newcount\nspan%
\newcount\ntemp%
\newdimen\hdsize%
\newdimen\newhdsize%
\newdimen\parasize%
\newdimen\spreadwidth%
\newdimen\thicksize%
\newdimen\thinsize%
\newdimen\tablewidth%
\newif\ifcentertables%
\newif\ifendsize%
\newif\iffirstrow%
\newif\iftableinfo%
\newtoks\dbt%
\newtoks\hdtks%
\newtoks\savetks%
\newtoks\tableLETtokens%
\newtoks\tabletokens%
\newtoks\widthspec%
%
%
%
\tableinfotrue%
\catcode`\@=11
%
%
\def\tstrut{\vrule height3.1ex depth1.2ex width0pt}%
\def\and{\char`\&}
\def\tablerule{\noalign{\hrule height\thinsize depth0pt}}%
\thicksize=1.5pt
\thinsize=0.6pt
\def\thickrule{\noalign{\hrule height\thicksize depth0pt}}%
\def\ctr#1{\hfil\ #1\hfil}%
%
%
%
%
\tablewidth=-\maxdimen%
\spreadwidth=-\maxdimen%
\def\tabskipglue{0pt plus 1fil minus 1fil}%
%
%
\centertablestrue%
%
%
%
%
\parasize=4in%
\gdef\ARGS{########}
\gdef\headerARGS{####}
\def\@mpersand{&}
{\catcode`\|=13
\gdef\letbarzero{\let|0}
\gdef\letbartab{\def|{&&}}%
\gdef\letvbbar{\let\vb|}%
}
{\catcode`\&=4
\def\ampskip{&\omit\hfil&}
\catcode`\&=13
\let&0
\xdef\letampskip{\def&{\ampskip}}%
\gdef\letnovbamp{\let\novb&\let\tab&}
}
\def\begintable{
   \begingroup%
   \catcode`\|=13\letbartab\letvbbar%
   \catcode`\&=13\letampskip\letnovbamp%
   \def\multispan##1{
      \omit \mscount##1%
      \multiply\mscount\tw@\advance\mscount\m@ne%
      \loop\ifnum\mscount>\@ne \sp@n\repeat%
   }
   \def\|{%
      &\omit\widevline&%
   }%
   \ruledtable
}
\long\def\ruledtable#1\endtable{%
%
%
%
   \offinterlineskip
   \tabskip 0pt
   \def\widevline{\vrule width\thicksize}
   \def\endrow{\@mpersand\omit\hfil\crnorm\@mpersand}%
   \def\crthick{\@mpersand\crnorm\thickrule\@mpersand}%
   \def\crthickneg##1{\@mpersand\crnorm\thickrule
          \noalign{{\skip0=##1\vskip-\skip0}}\@mpersand}%
   \def\crnorule{\@mpersand\crnorm\@mpersand}%
   \def\crnoruleneg##1{\@mpersand\crnorm
          \noalign{{\skip0=##1\vskip-\skip0}}\@mpersand}%
   \let\nr=\crnorule
   \def\endtable{\@mpersand\crnorm\thickrule}%
   \let\crnorm=\cr
%
%
   \edef\cr{\@mpersand\crnorm\tablerule\@mpersand}%
   \def\crneg##1{\@mpersand\crnorm\tablerule
          \noalign{{\skip0=##1\vskip-\skip0}}\@mpersand}%
   \let\ctneg=\crthickneg
   \let\nrneg=\crnoruleneg
   \the\tableLETtokens
%
%
   \tabletokens={&#1}
%
%
   \countROWS\tabletokens\into\nrows%
   \countCOLS\tabletokens\into\ncols%
%
%
   \advance\ncols by -1%
   \divide\ncols by 2%
   \advance\nrows by 1%
%
%
   \iftableinfo %
      \immediate\write16{[Nrows=\the\nrows, Ncols=\the\ncols]}%
   \fi%
%
%
   \ifcentertables
      \ifhmode \par\fi
      \line{
      \hss
   \else %
      \hbox{%
   \fi
      \vbox{%
         \makePREAMBLE{\the\ncols}
         \edef\next{\preamble}
         \let\preamble=\next
         \makeTABLE{\preamble}{\tabletokens}
      }
      \ifcentertables \hss}\else }\fi
   \endgroup
   \tablewidth=-\maxdimen
   \spreadwidth=-\maxdimen
}
\def\makeTABLE#1#2{
   {
   \let\ifmath0
   \let\header0
   \let\multispan0
%
%
   \ncase=0%
   \ifdim\tablewidth>-\maxdimen \ncase=1\fi%
   \ifdim\spreadwidth>-\maxdimen \ncase=2\fi%
   \relax
%
   \ifcase\ncase %
      \widthspec={}%
   \or %
      \widthspec=\expandafter{\expandafter t\expandafter o%
                 \the\tablewidth}%
   \else %
      \widthspec=\expandafter{\expandafter s\expandafter p\expandafter r%
                 \expandafter e\expandafter a\expandafter d%
                 \the\spreadwidth}%
   \fi %
   \xdef\next{
      \halign\the\widthspec{%
      #1
      \noalign{\hrule height\thicksize depth0pt}
      \the#2\endtable
%
      }
   }
   }
   \next
}
\def\makePREAMBLE#1{
   \ncols=#1
   \begingroup
   \let\ARGS=0
   \edef\xtp{\widevline\ARGS\tabskip\tabskipglue%
   &\ctr{\ARGS}\tstrut}
   \advance\ncols by -1
   \loop
      \ifnum\ncols>0 %
      \advance\ncols by -1%
      \edef\xtp{\xtp&\vrule width\thinsize\ARGS&\ctr{\ARGS}}%
   \repeat
   \xdef\preamble{\xtp&\widevline\ARGS\tabskip0pt%
   \crnorm}
   \endgroup
}
\def\countROWS#1\into#2{
   \let\countREGISTER=#2%
   \countREGISTER=0%
   \expandafter\ROWcount\the#1\endcount%
}%
\def\ROWcount{%
   \afterassignment\subROWcount\let\next= %
}%
\def\subROWcount{%
   \ifx\next\endcount %
      \let\next=\relax%
   \else%
      \ncase=0%
      \ifx\next\cr %
         \global\advance\countREGISTER by 1%
         \ncase=0%
      \fi%
      \ifx\next\endrow %
         \global\advance\countREGISTER by 1%
         \ncase=0%
      \fi%
      \ifx\next\crthick %
         \global\advance\countREGISTER by 1%
         \ncase=0%
      \fi%
      \ifx\next\crnorule %
         \global\advance\countREGISTER by 1%
         \ncase=0%
      \fi%
      \ifx\next\crthickneg %
         \global\advance\countREGISTER by 1%
         \ncase=0%
      \fi%
      \ifx\next\crnoruleneg %
         \global\advance\countREGISTER by 1%
         \ncase=0%
      \fi%
      \ifx\next\crneg %
         \global\advance\countREGISTER by 1%
         \ncase=0%
      \fi%
      \ifx\next\header %
         \ncase=1%
      \fi%
      \relax%
      \ifcase\ncase %
         \let\next\ROWcount%
      \or %
         \let\next\argROWskip%
      \else %
      \fi%
   \fi%
   \next%
}
\def\counthdROWS#1\into#2{%
\dvr{10}%
   \let\countREGISTER=#2%
   \countREGISTER=0%
\dvr{11}%
\dvr{13}%
   \expandafter\hdROWcount\the#1\endcount%
\dvr{12}%
}%
\def\hdROWcount{%
   \afterassignment\subhdROWcount\let\next= %
}%
\def\subhdROWcount{%
   \ifx\next\endcount %
      \let\next=\relax%
   \else%
      \ncase=0%
      \ifx\next\cr %
         \global\advance\countREGISTER by 1%
         \ncase=0%
      \fi%
      \ifx\next\endrow %
         \global\advance\countREGISTER by 1%
         \ncase=0%
      \fi%
      \ifx\next\crthick %
         \global\advance\countREGISTER by 1%
         \ncase=0%
      \fi%
      \ifx\next\crnorule %
         \global\advance\countREGISTER by 1%
         \ncase=0%
      \fi%
      \ifx\next\header %
         \ncase=1%
      \fi%
\relax%
      \ifcase\ncase %
         \let\next\hdROWcount%
      \or%
         \let\next\arghdROWskip%
      \else %
      \fi%
   \fi%
   \next%
}%
{\catcode`\|=13\letbartab
\gdef\countCOLS#1\into#2{%
   \let\countREGISTER=#2%
   \global\countREGISTER=0%
   \global\multispancount=0%
   \global\firstrowtrue
   \expandafter\COLcount\the#1\endcount%
   \global\advance\countREGISTER by 3%
   \global\advance\countREGISTER by -\multispancount
}%
\gdef\COLcount{%
   \afterassignment\subCOLcount\let\next= %
}%
{\catcode`\&=13%
\gdef\subCOLcount{%
   \ifx\next\endcount %
      \let\next=\relax%
   \else%
      \ncase=0%
      \iffirstrow
         \ifx\next& %
            \global\advance\countREGISTER by 2%
            \ncase=0%
         \fi%
         \ifx\next\span %
            \global\advance\countREGISTER by 1%
            \ncase=0%
         \fi%
         \ifx\next| %
            \global\advance\countREGISTER by 2%
            \ncase=0%
         \fi
         \ifx\next\|
            \global\advance\countREGISTER by 2%
            \ncase=0%
         \fi
         \ifx\next\multispan
            \ncase=1%
            \global\advance\multispancount by 1%
         \fi
         \ifx\next\header
            \ncase=2%
         \fi
         \ifx\next\cr       \global\firstrowfalse \fi
         \ifx\next\endrow   \global\firstrowfalse \fi
         \ifx\next\crthick  \global\firstrowfalse \fi
         \ifx\next\crnorule \global\firstrowfalse \fi
         \ifx\next\crnoruleneg \global\firstrowfalse \fi
         \ifx\next\crthickneg  \global\firstrowfalse \fi
         \ifx\next\crneg       \global\firstrowfalse \fi
      \fi
\relax
      \ifcase\ncase %
         \let\next\COLcount%
      \or %
         \let\next\spancount%
      \or %
         \let\next\argCOLskip%
      \else %
      \fi %
   \fi%
   \next%
}%
\gdef\argROWskip#1{%
   \let\next\ROWcount \next%
}
\gdef\arghdROWskip#1{%
   \let\next\ROWcount \next%
}
\gdef\argCOLskip#1{%
   \let\next\COLcount \next%
}
}
}
\def\spancount#1{
   \nspan=#1\multiply\nspan by 2\advance\nspan by -1%
   \global\advance \countREGISTER by \nspan
   \let\next\COLcount \next}%
\def\dvr#1{\relax}%
\def\header#1{%
\dvr{1}{\let\cr=\@mpersand%
\hdtks={#1}%
\counthdROWS\hdtks\into\hdrows%
\advance\hdrows by 1%
\ifnum\hdrows=0 \hdrows=1 \fi%
\dvr{5}\makehdPREAMBLE{\the\hdrows}%
\dvr{6}\getHDdimen{#1}%
{\parindent=0pt\hsize=\hdsize{\let\ifmath0%
\xdef\next{\valign{\headerpreamble #1\crnorm}}}\dvr{7}\next\dvr{8}%
}%
}\dvr{2}}
\def\makehdPREAMBLE#1{
\dvr{3}%
\hdrows=#1
{
\let\headerARGS=0%
\let\cr=\crnorm%
\edef\xtp{\vfil\hfil\hbox{\headerARGS}\hfil\vfil}%
\advance\hdrows by -1
\loop
\ifnum\hdrows>0%
\advance\hdrows by -1%
\edef\xtp{\xtp&\vfil\hfil\hbox{\headerARGS}\hfil\vfil}%
\repeat%
\xdef\headerpreamble{\xtp\crcr}%
}
\dvr{4}}
\def\getHDdimen#1{%
\hdsize=0pt%
\getsize#1\cr\end\cr%
}
\def\getsize#1\cr{%
\endsizefalse\savetks={#1}%
\expandafter\lookend\the\savetks\cr%
\relax \ifendsize \let\next\relax \else%
\setbox\hdbox=\hbox{#1}\newhdsize=1.0\wd\hdbox%
\ifdim\newhdsize>\hdsize \hdsize=\newhdsize \fi%
\let\next\getsize \fi%
\next%
}%
\def\lookend{\afterassignment\sublookend\let\looknext= }%
\def\sublookend{\relax%
\ifx\looknext\cr %
\let\looknext\relax \else %
   \relax
   \ifx\looknext\end \global\endsizetrue \fi%
   \let\looknext=\lookend%
    \fi \looknext%
}%
%
%
\def\tablelet#1{%
   \tableLETtokens=\expandafter{\the\tableLETtokens #1}%
}%
\catcode`\@=12
%

\magnification=1200
\oneandahalfspace
\tolerance=5000

\def\ds{\displaystyle}
\def\npb{Nucl. Phys. B}
\def\plb{Phys. Lett. B}
\def\mc{{\cal M}}
\def\hc{{\cal H}}
\def\gc{{\cal G}}
\def\dc{{\cal D}}
\def\fc{{\cal F}}
\def\tc{{\cal T}}
\def\hm{ \widehat {{\cal H}} }
\def\hu{ h_{1,1}}
\def\hd{ h_{2,1}}
\def\Z{ \hbox{\bf Z}}
\def\N{ \hbox{\bf N}}
\def\lb{\lbrack}
\def\rb{\rbrack}
\def\crs{\cr\noalign{\smallskip}}
\def\crm{\cr\noalign{\medskip}}

\footnote{}{\singlespace $^*$ Internet address:
cscpro7@caracas1.vnet.ibm.com}
\rightline{UCVFC/DF-1-92}
\title {PERIODS AND DUALITY SYMMETRIES IN CALABI-YAU COMPACTIFICATIONS}
\author {Anamar\'{\i}a  Font $^*$}
\affil {Departamento  de F\'{\i}sica, Facultad de Ciencias,
        Universidad Central de Venezuela,
        A.P.20513, Caracas 1020-A, Venezuela}

\abstract
We derive the period structure of several one-modulus Calabi-Yau
manifolds. With this knowledge we then obtain the generators
of the duality group and the mirror map that defines the
physical variable $t$ representing the radius of compactification.
We also describe the fundamental region of $t$ and
discuss its relation with automorphic functions.
As a byproduct of our analysis we compute the non-perturbative
corrections of Yukawa couplings.

\endtitlepage

\leftline{\bf 1. Introduction}\smallskip
Different string vacua can be related by continuous deformations
of some free parameters called moduli. From the compactification
point of view moduli deformations can be roughly interpreted as
changes in the metric of the internal manifold. In the space-time
effective theory moduli correspond to massless scalars with flat
potential whose non-zero VEVs signal the continuous deformations.
The space of moduli of a given string vacuum is generically symmetric
under certain transformations such as the $R \leftrightarrow 1/R$
duality of circle compactifications. Duality symmetries are an
important tool in the study of string vacua. For instance, they can
be used to infer how non-perturbative effects could modify the
low-energy effective theory $\/eff/$.
\par
The moduli space of Calabi-Yau (CY) compactifications has been
studied by several authors $\/moduli/$.
In CY threefolds moduli associated to zero modes
of the metric are directly linked to $(1,1)$- and $(2,1)$-harmonic
forms. Moduli of type $(1,1)$ and $(2,1)$ correspond respectively to
deformations of the K{\"a}hler form and the complex structure. A
remarkable feature of the moduli space is its factorization into
$\mc_{(1,1)} \times \mc_{(2,1)}$ where
$\mc_{(1,1)}$ and $\mc_{(2,1)}$ are special K\"ahler manifolds
of dimension given respectively by the Hodge numbers $\hu$ and
$\hd$. $(2,1)$-Moduli have a geometrical meaning in terms of periods
of the holomorphic 3-form and it can be shown that their duality
symmetries are described by a subgroup of $Sp(2\hd + 2, \Z)$.
\par
On the other hand, $(1,1)$-moduli do not have such
a geometrical counterpart and their duality symmetries are rather
stringy in character. However, as a consequence of the mirror
symmetry $\/espejo/$ that exchanges the
r\^ole of $(1,1)$- and $(2,1)$-moduli
it follows that the $(1,1)$-duality is a subgroup of
$Sp(2\hu + 2, \Z)$. In practice we are mostly interested in the
$(1,1)$-moduli. We know that a ``breathing mode" associated to
the radius of the internal manifold is always present. Determining
the symmetries acting on the corresponding massless field is of
the utmost relevance in the analysis of the effective theory.
\par
At present it is not known how to determine systematically the
proper subgroup of
$Sp(2\hu + 2, \Z) \times  Sp(2\hd + 2, \Z)$ that acts as duality
symmetries of the moduli space of a CY threefold. In a seminal
work, Candelas {\it et al} $\/CDGP/$ found the particular
$Sp(2\hu + 2, \Z)$ generators in an specific model. In this note
we wish to extend their methods to the analysis of several examples.
Our results are interesting in themselves as they can be used to
study the effective theory of strings compactified on CY manifolds.
They also constitute a further step towards more general
developments.
\par
This note is organized as follows. In section 2 we introduce
the models we will consider, find their mirror partners and
describe some basic properties of their moduli space.
In section 3 we review briefly the
relation between periods and duality symmetries.
In section 4 we derive the differential equations
satisfied by the periods of the mirror manifold and then
find explicit solutions. Using these results in section 5
we determine the corresponding $Sp(4,\Z)$ duality
generators. In section 6 we obtain the mirror maps for
our models, discuss their relation to automorphic forms
and compute the Yukawa couplings.
Conclusions are presented in section 7.
\bigskip

\leftline { \bf 2. Models}\smallskip
We will only consider CY threefolds with
$\hu = 1$.
Besides the quintic manifold $CP_4(5)$ studied in
\Ref{CDGP}  there
exist other relatively simple CY threefolds with
$\hu = 1$. These are defined as hypersurfaces $\hc$ in the
weighted projective space $WCP_4$. Our notation and the
corresponding defining polynomials are given below:
$$\eqalign{
A) \ \  WCP_4(2,1,1,1,1)_{-204} \ \ &: \ \
W_{0A} = X_1^3 + X_2^6 + X_3^6 + X_4^6 + X_5^6 = 0 \cr
B) \ \  WCP_4(1,1,1,1,4)_{-296} \ \ &: \ \
W_{0B} = X_1^8 + X_2^8 + X_3^8 + X_4^8 + X_5^2 = 0 \cr
C) \ \  WCP_4(2,1,1,1,5)_{-288} \ \ &: \ \
W_{0C} = X_1^5 + X_2^{10} + X_3^{10} + X_4^{10} + X_5^2 = 0 \cr
 } \eqno(wcero)$$
The numbers inside parentheses refer to the weights $n_m$ of
the $X_m$ coordinates. Notice that the $W_0$ are quasihomogeneous
functions of degree $d=(6,8,10)$ for models $(A,B,C)$,
e.g. $W_{0A}(\lambda^{n_m} X_m) = \lambda^6 W_{0A}(X_m)$.
These manifolds were first discussed in \Ref{SW} where their Euler
characteristic (shown above as a subscript) was also computed.
\par
Due to the link $\/Landau/$ between $N=2$ superconformal theories and
renormalization group fixed points of Landau-Ginzburg superpotentials
the above CY manifolds can also be described in terms of tensor
products of $N=2$ minimal models with diagonal invariants. From
the given superpotentials $W_{0A}$, $W_{0B}$, $W_{0C}$, we identify
these tensor products respectively as
$A \equiv 1 \ 4^4 = (1,4,4,4,4)$,
$B \equiv  6^4 = (6,6,6,6)$,
$C \equiv 3 \ 8^3 = (3,8,8,8)$; where $(k_1, \cdots, k_r)$ are
the levels of the $N=2$ minimal theories. Since a quadratic term
corresponds to a trivial $k=0$ theory the last two models only involve
four factors.
\par
In the Gepner construction $\/Gepner87/$
of the tensor models the Hodge numbers
are calculated as the number of $27$ and $\overline {27}$ fields.
In the above cases the results are $\hu = 1$ and $\hd^A=103$,
$\hd^B=149$ and $\hd^C=145$ $\/Lutken/$. These models have then many
$(2,1)$-moduli which can in fact be understood as the coefficients
of monomials in the $X_m$ that can be added to deform the $W_0$
while preserving their quasihomogeneity of the given degree.
\par
The $(1,1)$-modulus of the above manifolds can be interpreted
as the radius of the spaces. Our goal is to find the duality
symmetries of this modulus. To this end we will consider the
associated mirror manifolds that have $\hat \hu = \hd$ and
$\hat \hd = \hu = 1$. We will then find the generators of duality
symmetries of the $(2,1)$-modulus of the mirror manifolds
by exploiting its
geometrical interpretation in terms of periods. In virtue of the
mirror operation $\/espejo/$
these symmetries translate into symmetries of
the $(1,1)$-modulus of the original manifold.
\par
To obtain the mirrors of our models we use its $N=2$ tensor product
description. A $(k_1, \cdots, k_r)$ minimal tensor model has a
large group of symmetries given by ${\cal S} = Z_{k_1+2} \times
\cdots  \times Z_{k_r+2}$. Dividing by subgroups of ${\cal S}$
leads to new models. In particular,
it has been argued $\/AL/$ that modding by the maximal
subgroup of ${\cal S}$ yields the mirror of the original model.
Discrete symmetries and moddings of $N=2$ tensor products have
been studied in detail in \Ref{FIQS} . We now review briefly
the basic case of modding by a $Z_M$ subgroup. $Z_M$ is generated
by a modding vector
$$\Gamma = (\gamma_1, \cdots, \gamma_r) \eqno(mvec)$$
satisfying the supersymmetry preserving condition
$$ \sum_{i=1}^r \ { {\gamma_i}\over {k_i + 2} } =
\hbox{integer} \eqno(susy)$$
The order $M$ is the least integer such that
$M(\gamma_1, \cdots, \gamma_r) = 0 \hbox{mod} \
(k_1 +2 , \cdots, k_r +2)$. In terms of the $X_m$ coordinate
fields this $Z_M$ modding acts as a phase transformation
$$(X_1, \cdots, X_r) \rightarrow (\sigma^{\gamma_1 n_1}X_1,
\cdots, \sigma^{\gamma_r n_r}X_r) \eqno(fase)$$
where $\sigma = e^{2\pi i/d}$.
A multiple modding by $ G = Z_{M_1} \times \cdots \times Z_{M_P}$
with $M_b$ divisible by $M_a$ for $b \geq a$ is also possible.
Each $Z_{M_a}$ is generated by an independent modding $\Gamma_a$.
\par
Dividing by $G$ implies projecting out states in the original
spectrum while introducing new twisted states. The resulting
spectrum is derived by incorporating these orbifold effects in the
Gepner construction. Given our $A,B,C$ models we then look
for moddings that produce a new spectrum with $\hat \hd = 1$.
In each case we find that such a result is achieved by a maximal
subgroup of ${\cal S}$. The corresponding symmetries and
modding generators are given by
$$\vbox{\halign to 9.0 cm{
\hfil $#$ \hfil & \hfil $#$ \hfil & \hfil $#$ \hfil \cr
 A) &  1 \ 4^4 & \Gamma_1=(1,5,0,0,5) \cr
 & G= Z_6\times Z_6 \times Z_6  & \Gamma_2=(1,0,5,0,5) \cr
 && \Gamma_3=(1,0,0,5,5) \cr\noalign{\medskip}
 B) &  6^4 &  \Gamma_1=(7,2,2,5) \cr
 & G= Z_8\times Z_8 \times Z_8  & \Gamma_2=(7,2,5,2) \cr
&& \Gamma_3=(7,5,2,2) \cr\noalign{\medskip}
 C) &  3 \ 8^3  & \Gamma_1=(0,4,3,3) \cr
 & G= Z_{10}\times Z_{10}  & \Gamma_2=(0,3,4,3) \cr} }
\eqno(mods)$$
{}From the resulting spectrum it is very simple to identify the
primary field associated to the lone $(2,1)$-modulus of the
mirror manifolds. In terms of the $X_m$ coordinate fields
these are given respectively by
$$\eqalign{
h_A &=-6 X_1X_2X_3X_4X_5 \cr
h_B &=-4 X_1^2X_2^2X_3^2X_4^2 \cr
h_C &=-5 X_1^2X_2^2X_3^2X_4^2 \cr} \eqno(pert)$$
The numerical factor is conventional.
One can check that these are the only monomials
(involving more than one $X_m$) of degree $d_A=6$, $d_B=8$,
$d_C=10$ that are invariant under the phase transformations
\(fase) with the $\Gamma$ generators specified in \(mods). In
cases $B$ and $C$, $\gamma_5=0$ effectively.
\par
In the CY approach the mirror manifolds are obtained by
dividing each space defined in \(wcero) by the respective group $G$.
The single $(2,1)$-modulus can be taken as the coefficient of
the only monomial in the $X_m$ that may be added to $W_0$ to give
a $G$-invariant polynomial of the given degree.
We now introduce a modulus $\psi$
and consider the family of hypersurfaces
$\hm$ defined by the perturbed polynomials
$$\eqalign{
A) \ \ \  W_A &= X_1^3 + X_2^6 + X_3^6 + X_4^6 + X_5^6 -
6\psi X_1X_2X_3X_4X_5 = 0 \cr
B) \ \ \  W_B &= X_1^8 + X_2^8 + X_3^8 + X_4^8 + X_5^2 -
4\psi X_1^2X_2^2X_3^2X_4^2 = 0 \cr
C) \ \ \  W_C &= X_1^5 + X_2^{10} + X_3^{10} + X_4^{10} + X_5^2 -
5\psi X_1^2X_2^2X_3^2X_4^2 = 0 \cr} \eqno(mirror)$$
$\psi$ parametrizes changes in the complex structure of the
family of mirror manifolds $\hm/G$.
\par
For future purposes we need to characterize the holomorphic 3-form
$\Omega$ of our mirror manifolds. For CY $(N-1)$-folds defined by
an equation $W=0$ in $WCP_N$, $\Omega$ can be explicitly
constructed as explained in \Ref{SW} . First introduce inhomogeneous
coordinates, e.g. $Y_i = X_i X_{N+1}^{-n_i/n_{N+1}}$, $i=1, \cdots, N$.
Upon this substitution in $W(\psi)$ the equation $W(\psi)=0$
becomes $\lb 1 + \widehat W(\psi) \rb = 0$, where
$\widehat W(\psi)$ is a polynomial in the $Y_i$. $\Omega$ can
be written as
$$ \Omega(\psi) = \rho(\psi) { {dY_1 \wedge \cdots \wedge dY_{N-1}}
 \over { {\partial \widehat W(\psi)}\over {\partial Y_N} } }
 \eqno(holo)$$
Here $\rho(\psi)$ takes into account
the freedom in the normalization of $\Omega$.
A particular $\rho(\psi)$ in fact represents a choice
of gauge for $\Omega$ $\/Stro/$. Notice also that condition \(susy)
guarantees $G$-invariance of $\Omega$.
\par
We have seen that
for our manifolds having $\hat \hd =1$ the $(2,1)$-moduli space
is just the complex 1-dimensional space of $\psi$'s. Some values
of $\psi$ are equivalent. Two points $\psi$ and $\psi'$
describe the same model if the change
$\psi \rightarrow \psi'$ in $W$ can be undone by performing
linear transformations of the $X_m$ coordinate fields. As
explained in \Ref{SW} transformations $X_m \rightarrow R_{mn}X_n$
that maintain the form of $W$ actually correspond to symmetries of
the space $\hm/G$. Hence, they must verify that for each
$g \in G$, $RgR^{-1}= g'$ for some $g' \in G$. It is easy to
check that the only transformations satisfying this condition
are field rescalings
$X_m \rightarrow \sigma^{q_m n_m} X_m$ with $q_m \in \Z$ and
$\sigma^d = 1$. This transformation clearly leaves $W_0$ invariant
while multiplying the perturbations $h$ by a phase that can be
reabsorbed in $\psi$. In this way we obtain the symmetries
$$ \psi \rightarrow \alpha\psi \ \ ; \ \ \alpha^p=1 \eqno(asim)$$
where $p=(6,4,5)$ for models $(A,B,C)$.
\par
Another important feature of the space of $\psi$'s is the existence
of singular points, i.e. values of $\psi$ for which the manifolds
defined by \(mirror) become singular. For instance, when
$ \psi \rightarrow \infty$ these manifolds degenerate into
``pinched" varieties such as that described by
$ W_A \rightarrow  X_1X_2X_3X_4X_5 = 0$. Other singular values
of $\psi$ appear when the holomorphic 3-form fails to be well
defined. This occurs when the derivatives $\partial W/\partial X_m$
all vanish simultaneously. In our examples we find that this
condition is satisfied only when $\psi$ takes values so that
$$\eqalign{
A) \ \  4\psi^6 &=1\cr
B) \ \  \psi^4 &=1\cr
C) \ \  4\psi^5 &=1 \cr}\eqno(spu)$$
Due to the symmetries in \(asim) all solutions for $\psi$ are
identified. We can thus make the following specific choices of
singular points $\psi_0$
$$\eqalign{
A) \ \  \psi_0&=2^{-1/3} \cr
B) \ \  \psi_0&=1 \cr
C) \ \  \psi_0&=2^{-2/5}\cr} \eqno(spd)$$
In the next sections we shall see how these singular points turn
out to be very important in the study of modular symmetries.
\bigskip

\leftline { \bf 3. Periods and Symmetries }\smallskip
In this section we review some facts about symmetries of
moduli spaces of CY manifolds. Our aim is to establish notation
and to introduce some concepts that will be needed in our
subsequent analysis.
\par
The $(2,1)$-moduli space coincides
with the space of complex structures thus affording a
description in terms of periods of the holomorphic
3-form $\Omega$ $\/CD/$. For our mirror manifolds with
$\hat \hd=1$, $\Omega$ can be expanded in a basis
of harmonic 3-forms denoted by
$(\alpha_1,\alpha_2,\beta^1, \beta^2)$. Thus
$$ \Omega = z^1\alpha_1 + z^2\alpha_2 - \gc_1\beta^1
- \gc_2\beta^2 \eqno(oex)$$
The 3-cycles dual to the harmonic 3-forms are denoted
by $(A^1, A^2, B_1, B_2)$ and chosen canonically so that
$$ \int_{A^b} \alpha_a = - \int_{B_a} \beta^b =
\delta_a^b \eqno(norma)$$
with other integrals vanishing.
\par
The coefficients in the expansion \(oex) are interpreted as
periods since they are obtained by integrating $\Omega$ over
the canonical 3-cycles. For future convenience we introduce
the period vector $P$
$$ P = \pmatrix { \gc_1 \crs \gc_2 \crs
z^1 \crs z^2 \crs}\eqno(pvec)$$
Notice that $P$ is a function of $\psi$.
The entries of $P$ are not all independent. In fact, it has been
shown $\/Gri1/$ that $\gc_a = \partial \gc/\partial z^a$, where the
prepotential $\gc(z^1,z^2)$ is a homogeneous function of degree
two. Hence, $z^1$ and $z^2$ are only defined projectively, their
ratio giving just one independent degree of freedom related to the
single $(2,1)$-modulus $\psi$.
\par
In the previous section we explained how the points $\psi$ and
$\alpha \psi$, $\alpha^p=1$, define the same model. In terms of
the period vector the symmetry $\psi \to \alpha\psi$ is
represented by some matrix $S$, this is
$$ P \to SP \ \ \ ; \ \ \ \psi \to \alpha\psi \eqno(smat)$$
This transformation of $P$ necessarily corresponds to a change
of homology basis and hence $S \in Sp(4,\Z)$. Other symmetries
of $P$ are related to monodromy properties of the periods. We
will see later that $P(\psi)$ is multivalued about the singular points
of $\psi$ discussed in section 1. Transport about these points
generates a transformation of $P$
$$ P \to T_{\psi_0}P \ \ \ ; \ \ \
(\psi-\psi_0) \to e^{2\pi i}(\psi -\psi_0)  \eqno(tmat)$$
This monodromy must reflect a monodromy of the integral homology
basis so that $T_{\psi_0} \equiv T \in Sp(4,\Z)$.
\par
Since other solutions of \(spu) are related to $\psi_0$ by
$ \psi_0 \to \alpha\psi_0$ we deduce that transport about the
singular points $\psi_l = \alpha^l \psi_0$, $l=1, \cdots , p-1$
is generated by a composition of $S$ and $T$, namely
$T_l = S^{-l} T S^l$. Transport about $ \psi \to \infty$ is
also obtained from $S$ and $T$. This follows because a circuit
enclosing $\infty$ and the remaining multi-valued points can
be deformed into a cycle enclosing no singularities.
In conclusion, the matrices $S$ and $T$ generate a group
$\dc \in Sp(4,\Z)$ of symmetries of $P$. We will refer to $\dc$
as the modular or duality group.
\par
So far we have only discussed the $(2,1)$-moduli space of the
mirror manifold and how symmetries of $\psi$ imply
symmetries of $P$. However, our main goal is to determine the
duality symmetries of the $(1,1)$-moduli space of the original
manifold. The link between the two spaces is provided by the
mirror operation $\/espejo/$.
Indeed, a vector analogous to $P$ can be introduced
for the original manifold with $\hu=1$. This follows from the
existence of a prepotential $\fc$ that gives the metric of the
special K\"ahler manifold $\mc_{(1,1)}$. $\fc$ is a homogeneous
function of degree two depending on two variables $w^1$, $w^2$,
defined only projectively. The ratio $w^1/w^2$ corresponds to the
$(1,1)$-modulus of the original manifold denoted by $t$.
We then define the vector
$$ \Pi = \pmatrix { \fc_1 \crs \fc_2 \crs w^1 \crs w^2 \crs}
\eqno(fvec)$$
where $\fc_a = \partial \fc / \partial w^a$. The mirror hypothesis
basically states that $P$ and $\Pi$ are equal up to an $Sp(4,\Z)$
transformation $\/AL/$. As a consequence, duality symetries of $P$
translate into duality symmetries of $\Pi$.
This is an important observation since a priori $\fc$, and thus $\Pi$,
are only known classically, i.e. without including quantum corrections.
Thus, to obtain the duality group of the $(1,1)$-modulus
of the original manifold it is enough to first determine the
period vector $P$ of the mirror manifold
and then derive its transformation properties
defined in eqs. \(smat) and \(tmat).
\par
In principle, the periods can be found by integrating
the explicit form of $\Omega$ given in \(holo) over a basis
of 3-cycles. The difficulty with this approach, partially followed
in \Ref{CDGP}, resides in the identification of canonical
3-cycles as well as in the actual evaluation of the integrals.
Fortunately, an alternative route, more suitable to generalizations,
can be taken. As noticed in $\/CDGP/$ the periods of $\Omega$
obey a differential equation whose knowledge simplifies the
analysis considerably. This differential equation, known as the
Picard-Fuchs equation, in fact follows from general results
in algebraic geometry $[\cite{Dwo},\cite{Gri2}]$.
These results have been reviewed
recently in the Physics and Mathematics literature in connection
both with CY manifolds and $N=2$ topological
theories $[\cite{VW}, \cite{BV} , \cite{CF} , \cite{AM} ,
\cite{Morri1}, \cite{Morri}, \cite{LSW}, \cite{c3}] $.
These issues will be the subject of the next section.
\bigskip

\leftline { \bf 4. Picard-Fuchs Equations and Solutions}\smallskip
In this section we will obtain the Picard-Fuchs (PF) equation
satisfied by the periods of $\Omega$ in our models $A,B,C$.
We will start by a short review of a method originally due to
Dwork $\/Dwo/$ and recently discussed by Cadavid and Ferrara
$\/CF/$. We have included the necessary generalizations
to the quasihomogeneous case.
\par
We will consider CY $(N-1)$-folds with a single
$(2,1)$-modulus $\psi$
and defined as a quotient space $\hm/G$. Here $\hm$ is a
hypersurface in $WCP_N(n_1, \cdots, n_{N+1})$, described by the
equation
$$W(X_m,\psi) = W_0(X_m) + \psi h(X_m) = 0 \eqno(wxp)$$
with $W(X_m,\psi)$ a homogeneous function of degree $d$, i.e.
$W(\lambda^{n_m} X_m, \psi) = \lambda^d W(X_m,\psi)$. $G$ is a
group of phase symmetries whose elements are specified by
modding vectors $\Gamma = (\gamma_1, \cdots , \gamma_{N+1})$
and whose action on the $X_m$ coordinates is given in
\(fase). The holomorphic $(N-1)$-form can be calculated from \(holo).
We denote the periods of $\Omega$ by $\omega_a$,
$a = 1, \cdots, b_{N-1}$, where
$b_{N-1}$ is the $(N-1)^{th}$-Betti
number ($b_1 =2$ for onefolds and $b_3=2\hd + 2= 4$
for threefolds with $\hd=1$).
The $\omega_a$ turn out to be the independent
solutions of the matrix equation
$$ { {dR(\psi)}\over {d\psi} } = R(\psi) M(\psi) \eqno(rmmat)$$
where $R$ and $M$ are $b_{N-1} \times b_{N-1}$ matrices.
\par
Matrix $M$ above can be determined as follows. First introduce
the sets
$$\eqalign{
I &= \{ V = (v_0, v_1, \cdots , v_{N+1}) \ \
\vert  \ \  v_j \in \N  \ \ ; \ \
dv_0 = \sum_{m=1}^{N+1} n_m v_m \} \crs
\tilde I &= \{ V \in I  \ \ ; \ \
0 < v_m < d/n_m  \ \ ; \ \ m=1, \cdots, N+1 \}
\cr}\eqno(isets)$$
The action of $G$ further constrains the elements of both sets
by requiring
$$ \sum_{m=1}^{N+1} n_m v_m \gamma_m  = 0 \ \hbox{mod}\ d
\ \ \ ; \ \ \   \forall \ \Gamma \in G \eqno(gcon)$$
The set $\tilde I$ defines $b_{N-1}$ fundamental monomials of the
form
$$\xi_a = X^{V_a} \ \ \ ; \ \ \
V_a \in \tilde I  \eqno(fmono)$$
where $X^V \equiv X_0^{v_0} X_1^{v_1} \cdots X_{N+1}^{v_{N+1}}$.
The next step is to define the covariant derivatives
$$D_m = X_m { {\partial}\over {\partial X_m} } +
X_0 X_m { {\partial W}\over {\partial X_m} }  \ \ \ ; \ \ \
 m=1, \cdots, N+1 \eqno(cder)$$
and to consider the quantity
$$Q_a = X_0 h(X) \xi_a \eqno(qa)$$
It can be shown that $Q_a$ can be expanded in terms
of fundamental monomials $\xi_b$ modulo covariant
derivatives $D_mX^U$, with $U \in I$, $u_0 < v_{a 0}$ $\/Dwo/$.
Hence we can write schematically
$$Q_a = M_{ab} \xi_b + (DX) \eqno(mmat)$$
$ M_{ab}$ is precisely the matrix that we are looking for.
Substituting in \(rmmat) leads to the PF equation for
$\omega_a$.
\par
To illustrate the application of the method just discussed
we will work out two examples in some detail. We will
consider first the simpler onefold described by
$$W = X_1^4 + X_2^4 + X_3^2 - 2 \psi X_1^2 X_2^2 = 0 \eqno(zc)$$
which corresponds to a $Z_4$ orbifold of a torus.
In this case set $\tilde I$ has two elements with
corresponding fundamental monomials
$$\eqalign{
\xi_1 &=X_0X_1X_2X_3 \crs
\xi_2 &=X_0^2 X_1^3 X_2^3 X_3 \cr} \eqno(bzc)$$
{}From \(zc) we recognize $h = -2 X_1^2 X_2^2 $. Acting with
$X_0 h$ on the $\xi$-basis yields
$$\eqalign{
Q_1 &=-2 X_0^2 X_1^3 X_2^3 X_3 \crs
Q_2 &=-2 X_0^3 X_1^5 X_2^5 X_3 \cr} \eqno(qzc)$$
$Q_1$ is already given as $Q_1 = -2\xi_2$. $Q_2$ needs to be
reduced. To this end consider the combination
$$\eqalign{
4D_1(X_0^2 X_1 X_2^5 X_3) &+ 4D_2(X_0^2 X_1^5 X_2 X_3)
+4\psi(D_1+D_2)\xi_2  - (D_1+D_2)\xi_1  \crs
&= 32(1-\psi^2) X_0^3 X_1^5 X_2^5 X_3 + 32\psi \xi_2
- 2 \xi_1 \cr}\eqno(combo)$$
Hence
$$ Q_2 = -{ 1\over {8(1-\psi^2)} }\xi_1
+ { {2\psi}\over {(1-\psi^2)} }\xi_2  + (DX) \eqno(qdos)$$
Matrix $M$ is then given by
$$M = \pmatrix{ 0 &  { {-1}\over {8(1-\psi^2)} }\crm
-2 &  { {2\psi}\over {(1-\psi^2)} }\cr}\eqno(mzc)$$
Therefore, the solution of \(rmmat) is of the form
$$R = \pmatrix{ \omega_1 & -{1\over 2}\omega_1^{\prime} \crs
 \omega_2 & -{1\over 2}\omega_2^{\prime} \cr} \eqno(rmat)$$
where $\omega_1$ and $\omega_2$ are independent solutions
of the PF equation
$$ { {d^2\omega}\over {d\psi^2} }
+ { {2\psi}\over {(\psi^2-1)} } { {d\omega}\over {d\psi} }
+ { {1}\over {4(\psi^2-1)} } \omega = 0 \eqno(pfzc)$$
We have also studied the onefold described by
$$W = X_1^3 + X_2^6 + X_3^2 - 3 \psi X_1 X_4^2 = 0 \eqno(zs)$$
After a similar analysis we arrive at the PF equation
$$ { {d^2\omega}\over {d\psi^2} }
+ { {12\psi^2}\over {(4\psi^3-1)} } { {d\omega}\over {d\psi} }
+ { {7\psi}\over {4(4\psi^3-1)} } \omega = 0 \eqno(pfzs)$$
This model corresponds to a $Z_6$ orbifold of a torus.
Equations \(pfzc) and \(pfzs) have been obtained in \Ref{c3}
in the context of $N=2$ topological theories.
\par
Let us now turn to threefolds and consider our model $C$.
In this case, account taken of constraint \(gcon), set
$\tilde I$ has four elements. The corresponding fundamental
monomials are
$$\eqalign{
\xi_1 &=X_0X_1X_2X_3X_4X_5 \crs
\xi_2 &=X_0^2 X_1^3 X_2^3 X_3^3 X_4^3 X_5 \crs
\xi_3 &=X_0^3 X_1^2 X_2^7 X_3^7 X_4^7 X_5 \crs
\xi_4 &=X_0^4 X_1^4 X_2^9 X_3^9 X_4^9 X_5 \cr
} \eqno(xmc)$$
Acting with $X_0 h$, $h = -5 X_1^2 X_2^2 X_3^2 X_4^2 $, gives
$$\eqalign{
Q_1 &=-5 X_0^2 X_1^3 X_2^3 X_3 ^3 X_4^3 X_5\crs
Q_2 &=-5 X_0^3 X_1^5 X_2^5 X_3^5 X_4^5 X_5 \crs
Q_3 &=-5 X_0^4 X_1^4 X_2^9 X_3^9 X_4^9 X_5 \crs
Q_4 &=-5 X_0^5 X_1^6 X_2^{11} X_3^{11} X_4^{11} X_5 \cr
} \eqno(qmc)$$
$Q_2$ and $Q_4$ need to be reduced. For $Q_2$ consider
$$D_1(X_0^2 X_2^5 X_3^5 X_4^5 X_5) =
5 X_0^2 X_1^5 X_2^5 X_3^5 X_4^5 X_5
-10\psi X_0^3 X_1^2 X_2^7 X_3^7 X_4^7 X_5 \eqno(ccombo)$$
Hence
$$Q_2 = -10\psi \xi_3 + (DX) \eqno(qdc)$$
Reduction of $Q_4$ is cumbersome but straightforward. The
end result for matrix $M$ is
$$M= \pmatrix { 0 & 0 & 0 & { 1\over {1000(4\psi^5-1)} }\crm
-5 & 0 & 0 & { {-2\psi}\over {5(4\psi^5-1)} }\crm
0 & -10\psi & 0 & { {18\psi^3}\over {(4\psi^5-1)}}\crm
0 & 0 & -5 & { {-40\psi^4}\over {(4\psi^5-1)}}\cr} \eqno(cmmat)$$
Equation \(rmmat) is then solved by a matrix $R$ with elements
$$\eqalign{
R_{a1} &= \omega_a\crs
R_{a2} &= -{1\over 5}\omega_a^{\prime} \crs
R_{a3} &= {1\over {50\psi}}\omega_a^{\prime\prime} \crs
R_{a4} &= {1\over {250\psi^2}}\omega_a^{\prime\prime}
-{1\over {250\psi}}\omega_a^{\prime\prime\prime}\cr}\eqno(rcmat)$$
where the $\omega_a$ are the independent solutions of the
PF equation
$$ { {d^4\omega}\over {d\psi^4} }
+ {{2(16\psi^5+1)}\over {\psi(4\psi^5-1)}}
{ {d^3\omega}\over {d\psi^3}}
+ {{2(29\psi^5-1)}\over {\psi^2(4\psi^5-1)} }
{ {d^2\omega}\over {d\psi^2} }
+ { {20\psi^2}\over {(4\psi^5-1)} } { {d\omega}\over {d\psi} }
+ {{\psi}\over {4(4\psi^5-1)} } \omega = 0  \eqno(pfmc)$$
The PF equations for models $A$ and $B$ are obtained {\it mutatis
mutandis}. As for model $C$ in both cases the PF equation
takes the form
$$ { {d^4\omega}\over {d\psi^4} }+
\sum_{j=0}^{3} \ C_j(\psi) { {d^j\omega}\over {d\psi^j} } = 0
\eqno(pfg)$$
The explicit expressions of the coefficients $C_j(\psi)$ are
recorded in Table 1. Equivalent results have
been obtained in \Ref{Morri}.
\par
Let us now comment on some generic properties of the PF
equations of the various models. They are all Fuchsian
equations with the regular singular points located at
$\psi = 0, \infty, \alpha^l \psi_0$,
$l=0, \cdots, p-1$. All equations can be transformed into
generalized hypergeometric equations $\/Slater/$
upon the change of variables $\zeta = (\psi/\psi_0)^p$.
Recall that $p=(6,4,5)$ for models $(A,B,C)$.
To obtain solutions around
the singular points it is actually simpler to directly apply Frobenius
method to the equation in the variable $\psi$. Below we treat
model $C$ in some detail.
\par
The solutions of the PF equation \(pfmc) around $\psi=0$ are given
by
$$\eqalign{
\omega_j(\psi) &= \psi^j F({{2j+1}\over {10}}, {{2j+1}\over {10}},
{{2j+1}\over {10}}, {{2j+1}\over {10}};
\overbrace { {{j+1}\over {5}}, {{j+2}\over {5}},
{{j+4}\over {5}}, {{j+5}\over {5}} } ; 4\psi^5) \crs
j&=0,1,3,4 \ \ \ \ ; \ \ \ \ \vert \psi \vert < \psi_0 \cr}
\eqno(zsmc)$$
The overbrace indicates that the entry equal to 1 must be dropped.
The generalized hypergeometric function
$F(a_1,a_2,a_3,a_4;c_1,c_2,c_3;\zeta)$ is defined as $\/Slater/$
$$F(a_1,a_2,a_3,a_4;c_1,c_2,c_3;\zeta) = \sum_{l=0}^{\infty} \
{ { (a_1)_l (a_2)_l (a_3)_l (a_4)_l }\over
{ (c_1)_l (c_2)_l (c_3)_l } } \  { {\zeta^l}\over {l !} }
\eqno(hgeo)$$
where $(a)_l = \Gamma(a + l)/\Gamma(a)$.
The solutions around $\psi=0$ for models $A$ and $B$ are given in
Table 2.
\par
The indicial equation about $\psi= \psi_0$ has two simple
roots $s=0,2$ and a double root $s=1$. We find three independent
analytic solutions (with exponents $s=0,1,2$) and one solution
with a logarithmic singularity (due to the repeated index $s=1$).
A simple expression for these solutions cannot be given but
for our purposes it is necessary to characterize the logarithmic
solution that can be written as
$$f(\psi) = g(\psi) \ln (\psi - \psi_0) + \hbox{analytic}
\eqno(fsol)$$
where $g(\psi)$ is itself a solution of the form
$$g(\psi) = \sum_{l=0}^{\infty} \  a_l
(\psi - \psi_0)^{l+1} \eqno(gsol)$$
with $a_1 = -a_0/4$. The rest of the coefficients follows
from recurrence relations obtained by substituting \(gsol) in
the PF equation \(pfmc). In models $A$ and $B$ the results
are analogous. There is just one logarithmic solution of
the form \(fsol) with $g(\psi)$ given by an expansion
of type \(gsol). In $A$, $a_1 = -5a_0/12$. In $B$,
$a_1 = -5a_0/16$. In all cases we can take $a_0=1$.
\par
The function $g(\psi)$ will bear heavily in our analysis
due to its relation to monodromy properties about $\psi=\psi_0$.
Since $f(\psi)$ is one of the independent solutions around
$\psi_0$ when continuing any solution $\omega$ near this point
there will be a piece proportional to $f$ plus analytic
terms, i.e. $\omega = \delta f + \hbox{analytic}$. We then conclude
that $\omega$ will transform as
$\omega \to \omega + 2\pi i \delta g$
under transvection about $\psi = \psi_0$.
\par
The indicial equation about $\psi= \infty$ has a quadruple root
$s=1/2$. Hence, three of the independent solutions have
logarithmic singularities. These solutions are found in a
standard way $\/Ince/$. They are expressed in terms of the
function
$$\eqalign{
y(s) &= { {2^{1/5}}\over {4\pi^2} } (4^{1/5} \psi)^{-s}
\sum_{l=0}^{\infty} \  {  {\Gamma(l + {s\over 5})
\Gamma(l + {{s+1}\over 5}) \Gamma(l + {{s+3}\over 5})
\Gamma(l + {{s+4}\over 5}) } \over
{ \Gamma^4(l + {{2s+9}\over {10}}) }  } \  { 1\over {(4\psi^5)^l} }
 \crm  &\vert \psi \vert > \psi_0
\ \ \ ; \ \ \ 0 \leq \arg \psi \leq {{2\pi}\over 5} \cr}\eqno(ys)$$
Then, a basis of independent solutions around $\psi=\infty$
is given by
$$y_i(\psi) = { {d^iy}\over {ds^i} }(s_0)
\ \ \ ; \ \ \  i=0,\cdots,3 \eqno(yinf)$$
where $s_0 = 1/2$.
This basis is characterized by its monodromy about $\psi=\infty$.
More precisely, under $(1/\psi) \to e^{2\pi i} (1/\psi)$
we have
$$\eqalign{
\tilde y_0 &\to \tilde y_0 \crs
\tilde y_1 &\to \tilde y_1  + 2\pi i \tilde y_0 \crs
\tilde y_2 &\to \tilde y_2  + 4\pi i \tilde y_1
- 4\pi^2 \tilde y_0 \crs
\tilde y_3 &\to \tilde y_3  + 6\pi i \tilde y_2
- 12\pi^2 \tilde y_1  - 8i\pi^3 \tilde y_0  \crs }\eqno(imono)$$
where we have defined $\tilde y_i = \psi^{s_0} y_i$. For future
purposes we need the explicit form of $y_0$ and $y_1$. Results
for all models are recorded in Table 3.
In this table $\Psi$ is the Digamma function. To arrive at
the expansions for $y_1$ use has been made of Gauss's multiplication
formula for the $\Gamma$-function. In all cases the region
of convergence is $\vert \psi \vert > \psi_0
\ ; \ 0 \leq \arg \psi \leq 2\pi/p $.
\par
We will also need the analytic continuation of the solution
$y_0$ to $\vert \psi \vert < \psi_0$. $y_0$ can be continued
as explained in $\/CDGP/$.
First its series expansion is converted into a contour
integral (a so-called Barnes integral $\/Slater/$)
and then the contour is deformed adequately.
The end result can be written as
$$y_0(\psi) = -
\sum_{n=1}^{\infty} \  \lambda(n) \beta(n) \psi^{s_0(n-1)}
\ \ \ ; \ \ \  \vert \psi \vert < \psi_0
 \eqno(anau)$$
where $d$ is the degree of quasihomogeneity and $s_0=(1,1/2,1/2)$
for models $(A,B,C)$. The coefficients
$\lambda(n)$ and $\beta(n)$ are given by
$$\eqalign{
A) \ \ \ \lambda(n) &= e^{5i\pi n/6}
\sin^3 {{\pi n}\over 6} \sin {{\pi n}\over 3} \crs
\beta(n) &= { {6^n}\over {6 \pi^4 (n-1) !} }
\Gamma^4 ({n\over 6}) \Gamma ({n\over 3}) \crm
B) \ \ \ \lambda(n) &= e^{7i\pi n/8}
\sin^3 {{\pi n}\over 8} \sin {{\pi n}\over 2} \crs
\beta(n) &= { {4^n}\over {8\pi^4 (n-1) !} }
\Gamma^4 ({n\over 8})  \Gamma ({n\over 2}) \crm
C) \ \ \ \lambda(n) &= e^{9i\pi n/10}
\sin^2 {{\pi n}\over {10}} \sin {{\pi n}\over 5}
 \sin {{\pi n}\over 2} \crs
\beta(n) &= { {20^{n/2}}\over {10\pi^4 (n-1) !} }
\Gamma^3 ({n\over {10}})
 \Gamma ({n\over 5}) \Gamma ({n\over 2}) \cr }
\eqno(bls)$$
Clearly, the form of $\lambda(n)$ restricts the values of $n$.
For instance, in case $C$, $n=(2j+1)\ \hbox{mod} \ 10$ with
$j=0,1,3,4$. We then recover the expected result that for
$\vert \psi \vert < \psi_0$, $y_0$ can be written as a combination
of the solutions around $\psi=0$ given in Table 1,
specifically
$$y_0(\psi) = -
\sum_{j} \  \lambda_j \beta_j \omega_j(\psi)
\ \ \ ; \ \ \  \vert \psi \vert < \psi_0  \eqno(anad)$$
where $\lambda_j \equiv\lambda ({j\over {s_0}} + 1)$ and
$\beta_j \equiv \beta ({j\over {s_0}} + 1)$
are obtained from \(bls).
\par
Another piece of information required is the monodromy
about $\psi=\psi_0$ of the functions
$$e_l(\psi) \equiv y_0(\alpha^l\psi) =
-\sum_{j} \  \lambda_j \beta_j \omega_j(\alpha^l\psi)
\ \ \ ; \ \ \  \vert \psi \vert < \psi_0  \eqno(ebase)$$
where $\alpha^p=1$. As explained before,
around $\psi=\psi_0$ necessarily
$$e_l(\psi) = \delta_l g(\psi) \ln (\psi - \psi_0) + \hbox{analytic}
\eqno(esol)$$
with $g(\psi)$ given in \(gsol). Therefore,
$$ e_l(\psi) \to e_l(\psi) + 2\pi i \delta_l g(\psi) \eqno(zmono)$$
under transvection about $\psi = \psi_0$.
The coefficients $\delta_l$ turn out to be
$$\eqalign{
A) \ \ \ \delta_l &= { {4\sqrt{3}}\over {\pi^2} }
\sum_j  \ \alpha^{j l} \lambda_j
\ \ \ ; \ \ \  j=0,1,3,4 \crm
B) \ \ \ \delta_l &= { {2\sqrt{2}}\over {\pi^2} }
\sum_j  \ \alpha^{j l} \lambda_j
\ \ \ ; \ \ \  j=0,1,2,3 \crm
C) \ \ \ \delta_l &= { 2\over {\pi^2} }
\sum_j  \ \alpha^{j l} \lambda_j
\ \ \ ; \ \ \  j=0,1,3,4 \cr } \eqno(emono)$$
To derive these results we proceed as in \Ref{CDGP}.
The starting point is the series expansion
of $e_l(\psi)$ obtained from \(anau). Expanding the
$\Gamma$-functions by means of Stirling's formula and taking
derivative we isolate the logarithmically divergent piece.
The coefficient of this piece is $\delta_l$ since in the
limit $\psi \to \psi_0$,
$ {{de_l}\over {d\psi}} \to \delta_l \ln (\psi - \psi_0)
+ \ \hbox{finite}$.
\par
The sums in \(emono) are easily evaluated. For $l=0, \cdots, p-1$
we obtain
$$\eqalign{
A) \ \ \ \delta_l &= { {3\sqrt{3}}\over {2\pi^2} }
 \alpha^{-l} (1,1,-3,2,2,-3)
\ \ \ ; \ \ \  \alpha= e^{2\pi i/6} \crm
B) \ \ \ \delta_l &= { {\sqrt{2}}\over {\pi^2} }
 \alpha^{-l/2} (1,1,-3,3)
\ \ \ ; \ \ \  \alpha= e^{2\pi i/4} \crm
C) \ \ \ \delta_l &= { 5\over {4\pi^2} }
 \alpha^{-3l} (1,-1,-2,0,2)
\ \ \ ; \ \ \  \alpha= e^{2\pi i/5} \cr} \eqno(nmono)$$
The usefulness of these results will be appreciated in
the next section.
\bigskip

\leftline { \bf 5. Duality Generators}\smallskip
In this section we will explain how the duality group is obtained
starting from the solutions $\omega_a(\psi)$ of the PF
equation. We will see that the problems of determining the
canonical periods defining $P(\psi)$ in terms of the
$\omega_a(\psi)$ and finding the duality generators are solved
simultaneously.
\par
The basic idea is to work with a basis of solutions of the PF
equation such that the phase transformation $\psi \to \alpha\psi$
and the transvection about $\psi= \psi_0$ are realized in a
simple way. To express $P(\psi)$ in this basis we will exploit
the fact that it must transform according to
\(smat) and \(tmat), with $S,T \in Sp(4,\Z)$.
Implementing $\psi \to \alpha \psi$ is best achieved by working
with solutions around $\psi=0$ that are defined without restrictions
in $\arg \psi$. Our basis will then be expressed in terms of the
$\omega_j$ given in Table 2. In fact, notice that we have already
introduced functions $e_l(\psi)$ that transform nicely under
$\psi \to \alpha \psi$, namely $e_l(\psi) \to e_{l+1}(\psi)$.
However, these functions are not quite adequate to our needs.
If we choose $e_l(\psi), l= 0, \cdots, 3$ as our basis we will
have to expand $e_4(\psi)$ in this basis and in general the
coefficients will not be integers. Moreover, the $\delta_l$
factors related to the monodromy matrix $T$ are not even real.
Fortunately, both these problems have a common cure.
\par
At this point we recall that we still have a gauge freedom
associated to the normalization of $\Omega(\psi)$. In practice
this means that we can express $P(\psi)$ in terms of periods
$$\hat \omega_j(\psi) =\rho(\psi) \omega_j(\psi) \eqno(ohat)$$
where $\rho(\psi)$ is a gauge fixing function. In particular,
we may try a new basis $\hat e_l(\psi)$ obtained from
the analytic continuation of
$\hat y_0(\alpha^l\psi)$, with $\hat y_0(\psi) =\rho(\psi) y_0(\psi)$,
and choose $\rho(\psi)$ so that the monodromy cofficients of the
$\hat e_l(\psi)$ are real. From \(nmono) we easily find the
appropriate gauge satisfying this requirement, namely
$$\rho(\psi) = (\psi, \psi^{1/2}, \psi^3) \eqno(gauge)$$
for models $(A,B,C)$. We will then choose a basis
$$\hat e_l(\psi) = -\left ( { {2\pi i}\over p} \right )^3
\sum_j \ \lambda_j \beta_j
\hat \omega_j(\alpha^l\psi) \eqno(tbasis)$$
for $l= 0, \cdots, 3$. The numerical factor is conventional.
The monodromy about $\psi=\psi_0$ of this new basis is easily
found from previous results. We obtain
$$\hat e_l(\psi) \to
\hat e_l(\psi) + \hat \delta_l \hat g(\psi) \eqno(tgee)$$
where
$$\eqalign{
A) \ \ \ \hat g(\psi) &= {{\pi^2}\over {3\sqrt{3}}}
\psi g(\psi) \crs
\hat \delta_l &=(1,1,-3,2) \crm
B) \ \ \ \hat g(\psi) &= {{\pi^2}\over {2\sqrt{2}}}
 \psi^{1/2} g(\psi) \crs
\hat \delta_l &=(1,1,-3,3) \crm
C) \ \ \ \hat g(\psi) &= {{4\pi^2}\over {25}} \psi^3 g(\psi) \crs
\hat \delta_l &=(1,-1,-2,0) \cr } \eqno(tmono)$$
Notice that $\hat g(\psi)$ is analytic at $\psi = \psi_0$.
\par
To determine the action on our basis of the transvection about
$\psi=\psi_0$ we need to write $\hat g(\psi)$
as a combination of the $\hat e_l(\psi)$. We obtain
$$\eqalign{
A) \ \ \ \hat g(\psi) &= \hat e_0(\psi) - \hat e_1(\psi) \crs
B) \ \ \ \hat g(\psi) &= \hat e_0(\psi) - \hat e_1(\psi) \crs
C) \ \ \ \hat g(\psi) &= \hat e_0(\psi) + \hat e_1(\psi)
\cr} \eqno(gtilde)$$
These results are found by using a trick explained in $\/CDGP/$.
First notice that the transformation \(tgee) is equivalent to
$$ \hat e_l(\psi) = { {\hat \delta_l}\over {2\pi i} }
\hat g(\psi) \ln (\psi - \psi_0) + \ \hbox{analytic}\eqno(ett)$$
Next, take $\psi=x$ real, $x > \psi_0$. Discontinuity
of $\ln (\psi - \psi_0)$ across the cut from $\psi_0$
to $\infty$ then implies in particular
$$ \hat \delta_1 \hat g(x) = \hat e_1(x-i\epsilon) -
\hat e_1(x + i\epsilon) \eqno(unot)$$
for $\epsilon$ infinitesimal. We now use
$$ \hat e_1(x-i\epsilon) =
\hat e_0(\alpha(x - i\epsilon)) $$
together with the expansion of
$\hat e_0(\psi) = \rho(\psi) y_0(\psi)$ for
$\vert \psi \vert > \psi_0$,
$0 \leq \arg \psi \leq {{2\pi}\over p}$, to arrive at
$$\hat e_0(\alpha(x - i\epsilon)) = \rho(\alpha) \alpha^{-s_0}
 \hat e_0(x+i\epsilon)$$
Substituting back in \(unot) and analytically continuing to
all $\psi$ lead to the final expressions in \(gtilde).
\par
To simplify the presentation we introduce a vector $E(\psi)$
defined by
$$ E = \pmatrix { \hat e_0 \crs \hat e_1 \crs
\hat e_2 \crs \hat e_3 \crs} \eqno(evec)$$
Under transvection about $\psi=\psi_0$ $E$ transforms as
$$ E \to T_E E \eqno(temat)$$
The matrix $T_E$ is completely determined from the above
results. We find
$$\eqalign{
A) \ \ \ T_E &= \pmatrix { 2 & -1 & 0 & 0 \crs
1 & 0 & 0 & 0 \crs
-3 & 3 & 1 & 0 \crs
2 & -2 & 0 & 1 \crs } \crm
B) \ \ \ T_E &= \pmatrix { 2 & -1 & 0 & 0 \crs
1 & 0 & 0 & 0 \crs
-3 & 3 & 1 & 0 \crs
3 & -3 & 0 & 1 \crs } \crm
C) \ \ \ T_E &= \pmatrix { 2 & 1 & 0 & 0 \crs
-1 & 0 & 0 & 0 \crs
-2 & -2 & 1 & 0 \crs
0 & 0 & 0 & 1 \crs } \cr } \eqno(teresu)$$
Notice that these matrices are not symplectic but have
determinant one.
\par
As advertised before, our new basis also affords an integral
realization of the phase symmetry $\psi \to \alpha\psi$. We
have $\hat e_l(\psi) \to \hat e_{l+1}(\psi)$ and for
$\hat e_4(\psi)$ we find
$$\eqalign{
A) \ \ \ \hat e_4(\psi) &= -\hat e_0(\psi) - \hat e_2(\psi) \crs
B) \ \ \ \hat e_4(\psi) &= -\hat e_0(\psi)  \crs
C) \ \ \ \hat e_4(\psi) &= -\hat e_0(\psi) - \hat e_1(\psi)
- \hat e_2(\psi) - \hat e_3(\psi) \cr } \eqno(ctran)$$
Then, under $\psi \to \alpha\psi$ the vector $E(\psi)$ transforms as
$$E \to S_E E \eqno(semat)$$
with the matrix $S_E$ given by
$$ S_E = \pmatrix { 0 & 1 & 0 & 0 \crs
0 & 0 & 1 & 0 \crs
0 & 0 & 0 & 1 \crs
\sigma_{30} & \sigma_{31} &
\sigma_{32} & \sigma_{33} \crs } \eqno(seresu)$$
The last row is read off from \(ctran).
\par
So far we have a basis in which transvection and phase
transformations are realized by integral matrices of
determinant one but not all symplectic. However, there
must exist a change of basis from $E(\psi)$ to $P(\psi)$
$$ P(\psi) = UE(\psi) \eqno(umat)$$
such that the matrices
$$ S= U S_E U^{-1} \ \ \ ; \ \ \
T = U T_E U^{-1} \eqno(ust)$$
are both integral and symplectic. To find this change of basis
we will first make some justified assumptions.
\par
Under transvection about $\psi = \psi_0$ the periods
necessarily transform into themselves plus a piece proportional
to $ \hat g(\psi)$. Moreover, the proportionality constant
is an integer (if it were not we could just introduce some
appropriate overall normalization factor). Hence,
by an $Sp(4, \Z)$ transformation we can always bring one of
the periods to be equal to $ \hat g(\psi)$.
For definiteness we choose $ z^2(\psi) = \hat g(\psi)$
so that the fourth row of $U$ follows from \(gtilde).
As remarked in $\/CDGP/$ geometrically $\hat g(\psi)$
corresponds to an
integral of $\Omega(\psi)$ around a cycle $A^2$ that is
unambiguously defined for all $\psi$ near $\psi_0$.
Likewise, we could choose another period equal to
$\hat e_0(\psi)$ since it transforms into itself
under transvection about $\psi = \infty$. To see
which period we could identify with $\hat e_0(\psi)$
notice that under transvection about $\psi= \psi_0$
it transforms as $\hat e_0 \to \hat e_0 + \hat g$,
meaning that $\hat e_0(\psi)$ is an
integral of $\Omega(\psi)$ around a cycle that necessarily
intersects $A^2$. Therefore, we take $\gc_2(\psi) =
\hat e_0(\psi)$.
In model $A$ we verified explicitly that integrating $\Omega(\psi)$
obtained from \(holo) along cycles $A^2, B_2$ defined as in
$\/CDGP/$ leads to periods $z^2(\psi) ,\  \gc_2(\psi)$
in complete agreement with our indirect results.
\par
It is very simple to check that with the above choices of
$z^2(\psi)$ and $\gc_2(\psi)$ the condition $T \in Sp(4,\Z)$
requires that the remaining periods
$z^1(\psi)$ and $\gc_1(\psi)$ be analytic at $\psi = \psi_0$.
Therefore, the monodromy matrix $T$ is given by
$$T = \pmatrix{ 1 & 0 & 0 & 0 \crs
0 & 1 & 0 & 1 \crs
0 & 0 & 1 & 0 \crs
0 & 0 & 0 & 1 \crs} \eqno(tresu)$$
in all three models. Geometrically, $z^1(\psi)$ and $\gc_1(\psi)$
correspond to integrals of $\Omega(\psi)$ around cycles remote
from the singular point at which all $\partial W/\partial X_m=0$
when $\psi = \psi_0$.
\par
Thus far we have determined two rows of the change of basis
matrix $U$. One of the eight unknown entries, say $U_{20}$,
can be taken to be zero due to an $Sp(4,\Z)$ freedom of
redefining $z^1(\psi)$ and $\gc_1(\psi)$ without altering
$z^2(\psi)$ and $\gc_2(\psi)$. Two of the remaining entries
are not independent but fixed by the form of $T$, i.e. by
requiring that $T \in Sp(4,\Z)$. The independent elements
are found by imposing that $S \in Sp(4,\Z)$. The end results
for $U$ and $S$ are given below
$$\eqalign{
A) \ \ \ U &= \pmatrix{ -{1\over 3} & {2\over 3} &
{1\over 3} & {1\over 3} \crs
1 & 0 & 0 & 0 \crs
-2 & 0 & 0 & 1 \crs
1 & -1 & 0 & 0 \crs} \crm
B) \ \ \ U &= \pmatrix{ -{1\over 2} & {1\over 2} &
{1\over 2} & {1\over 2} \crs
1 & 0 & 0 & 0 \crs
-3 & 0 & 1 & 2 \crs
1 & -1 & 0 & 0 \crs} \crm
C) \ \ \ U &= \pmatrix{ 1 & 1 & 0 & 1 \crs
1 & 0 & 0 & 0 \crs
2 & 0 & 1 & 1 \crs
1 & 1 & 0 & 0 \crs}
 \crs }
 \qquad; \qquad
\eqalign{
S &= \pmatrix{ 1 & -1 & 0 & 1 \crs
0 & 1 & 0 & -1 \crs
-3 & 0 & 1 & 0 \crs
-3 & 4 & 1 & -3 \crs } \crm
S &= \pmatrix{ 1 & -1 & 0 & 1 \crs
0 & 1 & 0 & -1 \crs
-2 & -2 & 1 & 2 \crs
-4 & 4 & 1 & -3 \crs } \crm
S &= \pmatrix{ -1 & -1 & 0 & 1 \crs
0 & -1 & 0 & 1 \crs
1 & 0 & -1 & 0 \crs
-1 & -3 & 1 & 2 \crs }
\crs } \eqno(usresu)$$
Actually, the above matrices are only determined up to the
freedom in redefining
$z^1(\psi)$ and $\gc_1(\psi)$ mentioned previously.
\par
As we explained in section 3, the matrices $S$ and $T$ generate
the duality group $\dc$ of the period vector $P(\psi)$. Notice
that $T$ is of infinite order whereas $S$ satisfies
$$\eqalign{
A) \ \ \ S^6 &= 1 \crs
B) \ \ \ S^8 &= 1 \crs
C) \ \ \ S^5 &= 1 \cr } \eqno(spot)$$
In example $B$, $S^4 = -1$ due to the branch point at $\psi=0$
introduced by the gauge $\rho(\psi) = \psi^{1/2}$. Transvection
about the singular points $\alpha^l \psi_0$ is given by
$$ T_l = S^{-l} T S^l \eqno(tlmat)$$
Transvection about $\psi=\infty$ is also computed from $S$ and $T$.
In all three models $T_{\infty}$ turns out to be
$$T_{\infty} = (ST)^{-p} \eqno(tinf)$$
For instance, in example $B$ we have
$$T_{\infty} T_3 T_2 T_1 T T^0 = 1$$
where $T^0 =-1$ is the transvection about $\psi=0$.
The above equation reflects the fact that a loop enclosing
all singularities can be deformed into a loop encircling
no singularities. Using \(tlmat) and $S^4=-1$ we indeed
find $T_{\infty} = (ST)^{-4}$. In the next section
we will need the explicit results for $(ST)^{-1}$
given below
$$\eqalign{
A) \ \ \ (ST)^{-1} &= \pmatrix{ 1 & 1 & 0 & 0 \crs
0 & 1 & 0 & 0 \crs
3 & 3 & 1 & 0 \crs
0 & -4 & -1 & 1 \crs} \crm
B) \ \ \ (ST)^{-1} &= \pmatrix{ 1 & 1 & 0 & 0 \crs
0 & 1 & 0 & 0 \crs
2 & 4 & 1 & 0 \crs
2 & -4 & -1 & 1 \crs} \crm
C) \ \ \ (ST)^{-1} &= \pmatrix{ -1 & 1 & 0 & 0 \crs
0 & -1 & 0 & 0 \crs
-1 & 1 & -1 & 0 \crs
0 & 3 & -1 & -1 \crs}
 \crs } \eqno(stmo)$$
$T_{\infty}$ is easily obtained using \(tinf).
\bigskip

\leftline{\bf 6. Mirror Maps, Automorphic Functions
and Yukawa Couplings}\smallskip
The ratio $t=w^1/w^2$ of the homogeneous coordinates of the
prepotential $\fc$ can be understood as the $(1,1)$-modulus
of the original manifold. The mirror map gives the relation between
$t$ and $\psi$. Our results in sections 4 and 5 allow a simple
derivation of $t$ as a function of $\psi$ as we will now explain.
\par
In $\/CDGP/$ $t$ was determined by first obtaining the explicit
$Sp(4, \Z)$ rotation relating $P$ and $\Pi$. It was also found
that the effect of $T_{\infty}$ on $t$ was an integer shift
by the order of the phase symmetry \(asim).
We may take this property as the definition of $t$. More precisely,
we will define $t$ so that under $(ST)^{-1}$ it transforms as
$$ t \to t +1 \eqno(itran)$$
In fact, imposing \(itran) does not fix $t$ completely. There is
a residual ambiguity that just reflects the above translational
symmetry. This prescription for finding $t(\psi)$ has also
been advocated in \Ref{Morri}.
\par
Thus, our strategy is to find the, necessarily integer, combinations
of the $z^a, \gc_a$ that give $w^1$ and $w^2$ such that their
ratio transforms as \(itran) under $(ST)^{-1}$. The appropriate
combinations, up to the ambiguity mentioned, are easily derived
from the results in \(stmo). We find
$$\eqalign{
A) \ \ \ t &= { {\gc_1}\over {\gc_2} } = { 1\over {3 \hat e_0} }
[ - \hat e_0 + 2 \hat e_1 + \hat e_2 + \hat e_3 ] \crm
B) \ \ \ t &= { {\gc_1}\over {\gc_2} } = { 1\over {2 \hat e_0} }
[  -\hat e_0 + \hat e_1 + \hat e_2 + \hat e_3 ] \crm
C) \ \ \ t &= -{ {\gc_1}\over {\gc_2} } = -{ 1\over { \hat e_0} }
[  \hat e_0 +  \hat e_1 + \hat e_3 ] \cr }\eqno(rresu)$$
The expansions of $t$ for $\vert \psi \vert < \psi_0$ follow
simply from \(tbasis) and the $\omega_j$ given in Table 2.
Notice that the freedom of adding a piece $m\gc_2$
to $\gc_1$ only produces an integer shift in $t$.
This would just imply that the fundamental domain
of $t$ is translated by $m$ in the Re $t$ direction
as allowed by the axionic symmetry \(itran).
\par
The expansions of $t$ for $\vert \psi \vert > \psi_0$ require
the analytic continuation of the periods. There is a simple
way of performing this continuation. We already know a basis
around $\psi=\infty$, namely
$\hat y_i(\psi) = \rho(\psi)y_i(\psi)$
with $y_i(\psi)$ given in \(yinf). Furthermore, the monodromy
of this basis about $\psi=\infty$ is also known. By comparing
$T_{\infty}$ computed from \(tinf) and \(stmo) with this
monodromy we can find how the periods are expanded in the
$\hat y$-basis. Clearly, $\gc_2 = ({{2\pi i}\over p})^3\hat y_0$
and for $\gc_1$
we find that it is proportional to $\hat y_1$ as expected
since $\hat y_1$ roughly transforms into itself plus $\hat y_0$
under transvection about $\psi = \infty$. Actually, we also
find that $\gc_1$ has a piece proportional to $\hat y_0$
that cannot be determined from the monodromy data only. However,
this piece is irrelevant since it can be cancelled against
a term proportional to $\gc_2$ that we could have added to
$\gc_1$ in \(rresu). In all examples the final result for $t$
can be written as
$$ t =  {p\over {2\pi i}} { {y_1}\over {y_0} } \eqno(tfinal)$$
where we have used $\hat y_1/\hat y_0= y_1 / y_0$, i.e. $t$
is gauge invariant.
Notice that $t$ transforms as $t \to t + p$ under $T_{\infty}$.
However, other generators such as $S$ or $T$ do not have
a simple action on $t$.
\par
The explicit expressions of $t(\psi)$ for $|\psi| > \psi_0$
are obtained from the expansions in Table 3. In all three
models, as well as in the quintic hypersurface studied in
$\/CDGP/$, the result takes the general form
$$t = {1\over {2\pi i}} \big\{ -\ln(c\psi)^p +
{1\over {\tilde y_0}} \sum_{l=0}^{\infty} \
{{(dl)!\ (c\psi)^{-pl}}\over {(n_1l)! \cdots (n_5l)!}}
[d\Psi(dl+1) - n_1\Psi(n_1l+1)- \cdots- n_5\Psi(n_5l+1)] \big\}
\eqno(funo)$$
where
$$\tilde y_0 = \sum_{l=0}^{\infty} \
{{(dl)!\ (c\psi)^{-pl}}\over {(n_1l)! \cdots (n_5l)!}}
\eqno(fdos)$$
Recall that $d$ is the degree of quasihomogeneity of the
defining polynomial, the $n_m$ are the weights of the $X_m$
coordinates and $p$ is the order of the phase symmetry
$\psi \to \alpha \psi$. Constant $c$ can be found from
$$(c\psi_0)^p = { {d^d}\over {n_1^{n_1} \cdots n_5^{n_5}} }
\eqno(ftres)$$
Then, $c=(6,16,20)$ in models $(A,B,C)$ as shown in Table 3.
Notice that the large $\psi$ limit of $t$ is
$$t \to -{p\over {2\pi i}} \ln(c\psi) \eqno(lpl)$$
since $d=n_1 + \cdots + n_5$.
\par
Knowing $t(\psi)$ we can determine the
fundamental region ${\cal T}$ of $t$ as the image of
the fundamental region of $\psi$ given by the wedge
$ 0 \leq \arg \psi \leq {{2\pi}\over p}$. Roughly
speaking, ${\cal T}$ consists of two adjacent triangles
with vertices at $t(\infty)=i\infty$, $t(\psi_0)$,
$t(0)$ and $[t(0) + 1]$ as shown schematically in
Figure 1. The vertices correspond to fixed points of the
duality group. $t(0)$ must be a fixed point of the $S$
generator since $\psi=0$ is fixed under the phase symmetry
$\psi \to \alpha \psi$. From \(rresu) and \(seresu) we
can verify that this is indeed the case.
On the other hand, $t(\psi_0)$ must be fixed under the
action of $T$ since this transvection leaves $\psi_0$ invariant.
To check this consider model $C$ for definiteness.
Under $T$ we have
$$t \to  - { {\gc_1}\over {\gc_2 + z^2} } \eqno(tt)$$
but $z^2(\psi_0) = 0$ and thus $t(\psi_0) \to t(\psi_0)$
as claimed. Notice also that $t(\infty)$ is fixed under
$T_{\infty}$.
\par
In all models $t$ has values at the vertices given by
$$\eqalign{
t(\psi_0) &= ia \crs
t(0) &= -{1\over 2} + ib \cr} \eqno(verts)$$
where $a$ and $b$ are constants with $ a > b$.
Constant $b$ is evaluated from the expansion of $t$
for $\vert \psi \vert < \psi_0$. We obtain
$$ b = {1\over 2} (\sqrt{3}, 1 + \sqrt{2},
\sqrt{ 5 + 2\sqrt{5} } ) \eqno(bvert)$$
for models $(A,B,C)$. For constant $a$, a rough numerical
computation gives
$$ a \sim (1.42, 1.70, 1.98) \eqno(avert)$$
These results follow from either the expansion of $t$
for $\vert \psi \vert > \psi_0$ or the expansion of $t$
for $\vert \psi \vert < \psi_0$ since they both converge as
$\psi \to \psi_0$.
\par
The boundary arc joining $t(\psi_0)$ to $t(0)$
is the image of the interval $[0,\psi_0]$. At $t(\psi_0)$
this arc is tangent to the imaginary axis as
expected from the fact that $t(0)$
is a fixed point of a generator of infinite order.
We can check this result
from the behavior of $t(\psi)$ near $\psi_0$ encoded
in the relation
$${{t(\psi) - t(\psi_0)}\over {(\psi - \psi_0)}} =
-i { {ap^3 \hat g_0}\over {(2\pi)^4 \hat y_0(\psi_0)} }
  \ln (\psi - \psi_0) \eqno(ttc)$$
 where $\hat g_0$ defined by $\hat g(\psi) =\hat g_0
 (\psi - \psi_0) + \cdots $ is found from \(tmono).
The above expression follows from \(rresu) and \(ett).
We can study the behavior of $t(\psi)$ near $\psi=0$ in
a similar manner. We find that at $t(0)$ the angle between
the arc and the imaginary axis is $\pi/p$ as it should
since $t(0)$ is fixed under a symmetry of order $p$.
\par
We now turn to a discussion of the relation of
$t$ to automorphic functions $\/Ford/$. To this purpose it
is instructive to study the simpler onefold models.
To obtain $t(\psi)$ in this case it is convenient
to adopt the approach taken
in $[\cite{VW} , \cite{LSW} , \cite{c3}] $.
Let us consider for example the $Z_6$ model. Defining
$t=\omega_1/\omega_2$, with $\omega_1$ and $\omega_2$
two independent solutions of the PF equation \(pfzs)
we find that $t$ satisfies
$$ \{ t, \psi \} = - { {\psi(20 \psi^3 - 41)}\over
{2(4\psi^3 - 1)^2} } \eqno(szs)$$
where $\{ f,x \} = { {f'''}\over {f'} } - {3\over 2}
\left ( { {f''}\over {f'} } \right )^2 $ is the Schwarzian
derivative. Introducing the variable
$$\kappa =  { {4 \psi^3}\over {4\psi^3 - 1} } \eqno(yuk)$$
then leads to
$$ \{ t, \kappa \} =  { 3\over {8(1 -\kappa)^2} } +
 { {23}\over {72\kappa (1 -\kappa)} } +  { 4 \over {9\kappa^2} }
\eqno(fri)$$
The solution of the above differential
equation is known to be the absolute
modular invariant of $PSL(2,\Z)$, i.e.
$\kappa(t)=J(t)$ $\/Klein/$. Hence, the mirror
map is implicitly given by
$$J(t) =  { {4 \psi^3}\over {4\psi^3 - 1} } \eqno(mapzs)$$
Recall that $J(t) = {1\over {1728}} \lb q^{-1} + 744 + 196884 q
+ \cdots \rb$, where $q = e^{2\pi i t}$ is the uniformizing variable.
{}From \(mapzs) we also conclude that the duality group
acting on $t$ is $PSL(2,\Z)$ as expected for a torus
compactification.
\par
In our threefold models $t$ should also represent the inverse
of an automorphic function of the duality group acting on
this variable. In general, automorphic functions
will be written as rational functions of
$\psi^p(t)$. A Fourier series of $\psi^p$ in powers of
$q = e^{2\pi i t}$ can be obtained from the expansion
of $t$ for large $\psi$. Below we give the first terms
of the series
$$\eqalign{
A) \ \ \ (6\psi)^6 &= {1\over q} + 2772 +
5703858\ q + 14332453152 \ q^2 + \cdots \crs
B) \ \ \ (16\psi)^4 &= {1\over q} + 15808 +
178476448\ q +  2473876932608 \ q^2  + \cdots  \crs
C) \ \ \ (20\psi)^5 &= {1\over q} + 179520 +
25399812000\ q +  4599352920320000\ q^2 + \cdots
\cr}  \eqno(glo)$$
$\psi^p$ is clearly invariant under $t \to t+1$. Being an
automorphic function, it must also be invariant under
the remaining duality generators.
Rather than invariant functions, for the purpose of
constructing duality invariant effective actions it
is necessary to know the equivalent of
modular forms, i.e. functions
that transform in an specific way under the duality
group. Of particular interest $\/eff/$ is the generalization
of the Dedekind cusp form $\eta(t)$.
\par
Let us now consider the computation of the
Yukawa coupling of the $(1,1)$-field. This coupling
is given by $\/CDGP/$
$$ \kappa_{ttt} = {1\over {\gc_2^2}}
\kappa_{\psi\psi\psi}
\left ( { {d\psi}\over {dt} } \right )^3 \eqno(yuco)$$
$\kappa_{\psi\psi\psi}$ is in turn given by
$$\kappa_{\psi\psi\psi} = W_3(\psi) \rho^2(\psi) \eqno(kapsi)$$
$W_3$ is the solution of the differential equation
$$ { {dW_3}\over {d\psi} } + {1\over 2} C_3(\psi)W_3 = 0
\eqno(ewt)$$
where $C_3(\psi)$ is the coefficient appearing in the
PF equation. We find
$$\eqalign{
A) \ \ \  \kappa_{\psi\psi\psi} &= c \ { {\psi^3}\over
{1 - 4\psi^6} } \crs
B) \ \ \  \kappa_{\psi\psi\psi} &= c \ { {\psi}\over
{1 - \psi^4} } \crs
C) \ \ \  \kappa_{\psi\psi\psi} &= c \ { {\psi^7}\over
{1 - 4\psi^5} } \cr } \eqno(karesu)$$
The normalization of $\kappa_{\psi\psi\psi}$ is fixed
by the condition that in the classical (large radius) limit
$$ \kappa_{ttt} \to  m_0 $$
where $m_0$ is the intersection number of the $(1,1)$-form
of the original hypersurface $\hc$ $\/SW/$.
For weighted projective spaces this topological invariant
can be shown to be equal to the degree of quasihomogeneity
divided by the product of the weights $\/m0/$.
Then, $m_0=(3,2,1)$ in models $(A,B,C)$.
\par
It was first conjectured $\/CDGP/$ and later proven $\/AM/$
that $\kappa_{ttt}$ can be expressed as
$$\kappa_{ttt} = m_0 + \sum_{k=1}^{\infty} \
{ { m_k k^3 q^k}\over {1 - q^k} } = m_0 + m_1q +
 (8m_2 + m_3)q^2 + \cdots \eqno(sinst)$$
where $m_k$ is the number of rational curves of degree $k$
in $\hc$. Given our expansions for $t(\psi)$ and
$\gc_2(\psi)$ it is straighforward to compute $\kappa_{ttt}$.
We have checked that its expansion is of the form \(sinst).
Table 4 shows the first coefficients $m_k$. Except for model
$C$ where we have taken the correct normalization $m_0=1$
the values obtained agree with recent results $\/Morri/$.
\par
To end this section we discuss briefly the structure of
the prepotential $\fc(t)$. From the explicit results for
$\Pi$ we can compute $\fc = {1\over 2} w^a \fc_a$ and
verify that it has the expected behavior
$$ \fc(t) = -{{m_0}\over 6} t^3 + f_2 t^2 + f_1 t +
\fc_{loop} + \hbox{non-perturbative} \eqno(prepo)$$
where the non-perturbative terms involve powers of
$e^{2\pi i t}$. As explained in \Ref{CDGP}, $\fc_{loop}$
is due to a sigma-model four-loop correction and its
form must be given by a universal constant times the
Euler characteristic $\chi$. Indeed, in all models
we find
$$\fc_{loop} =
 {{i\zeta(3)}\over {2(2\pi)^3} } \ \chi \eqno(univ)$$
 in agreement with the results of $\/CDGP/$.

\bigskip

\leftline { \bf 7. Conclusions}\smallskip
In this paper we have tackled the problem of finding
the duality symmetries of Calabi-Yau compactifications.
We focused our attention on the more tractable case of
one-modulus manifolds. In particular, we considered
three specific models and explained how the $Sp(4,\Z)$
duality generators could be found systematically.
\par
The starting point in our analysis was the Picard-Fuchs
differential equation satisfied by the periods of the
associated mirror manifolds. We obtained explicit solutions
of this equation that were then used to construct
simultaneously the symplectic basis of periods and
the duality generators. With these results we then
derived the relation between the $(1,1)$-modulus $t$
of the original manifold and the $(2,1)$-modulus $\psi$
of the mirror partner. Our prescription for computing the
mirror map $t(\psi)$ was based on the existence of the
axionic symmetry $t \to t +1$ together with the property
that $t$ can be written as a ratio of periods in the
symplectic basis. We found a general expression for $t(\psi)$
with parameters that follow directly from
the equation defining the mirror manifolds.
\par
$t$ is the interesting physical variable, its real and
imaginary parts correspond respectively to the antisymmetric
tensor field and the radius of compactification.
In general, the generators of the duality group act on $t$
in a complicated way. Nonetheless, knowing the mirror map
$t(\psi)$ allowed us to study the
basic features of the $t$ fundamental domain $\tc$.
{}From the shape of $\tc$ we deduce the existence of a
symmetry relating large and small radius. Equivalently,
we may say that there is a minimum value for the radius
of compactification given by Im $t(0)$. Other generic
properties of the fundamental region will likely hold
in more complicated models. For instance, the vertices
of $\tc$ are given by $t(0)$, $t(\psi_0)$ and $t(\infty)$
that correspond respectively to the Gepner point $\psi=0$
and the singular points $\psi=\psi_0$, $\psi=\infty$.
Furthermore, the internal angles of $\tc$ are determined
by the order of the symmetries that leave fixed these
special points.
\par
The results for $t(\psi)$
can be applied to determine other properties
of the models such as non-perturbative corrections to the
prepotentials and the Yukawa couplings. We computed the
first terms in the expansion of the Yukawa couplings.
{}From the Physics
point of view these corrections are needed in the
analysis of effective theories. They are also related to
the numbers of rational curves of the manifold $[\cite{CDGP},
\cite{AM}]$. For these numbers we obtained values in agreement
with recent work $\/Morri/$. We also verified that the
four-loop sigma-model correction to the prepotential
has a universal form.
\par
We remarked the fact that $t$ can be intrepreted as the
inverse of an automorphic function of the duality group
and $t(\psi)$ can be inverted to find the basic building
blocks of such functions.
The $Sp(4,\Z)$ duality generators that we have found in
principle can be used to construct modular forms following
a method developed in \Ref{FKLZ}.
It would be interesting to investigate
how the forms obtained in that approach are related
to the automorphic functions and Yukawa couplings derived
from the mirror maps. Work along these lines is in progress.

\bigskip\bigskip

\leftline { \bf Acknowledgements}\smallskip
I am deeply indebted to the Centro Cient\'{\i}fico IBM-Venezuela
for the use of its vital facilities. Hospitality at CERN,
LAPP and Institut de Physique-Univ. de Ne\^uchatel while
finishing this work is gratefully acknowledged. I have had
the benefit of useful conversations with L. Ib\'a\~nez,
L. Leal, F. Quevedo, R. Schimmrigk and F. Thuillier.
Very special thanks are due
to X. De La Ossa and P. Candelas for valuable remarks
and explanations and for kindly advising me on the use of
their codes to compute some of the results.

\vfill\eject


\references
\refis{CDGP} P. Candelas, X. De La Ossa, P. S. Green and L. Parkes,
\npb 359 (1991) 21; \plb 258 (1991) 118. \par
\refis{moduli}
S. Cecotti, S. Ferrara and L. Girardello, \plb 213 (1988) 443;
Int. J. Mod. Phys. A4 (1989) 2475; \hfil\break
S. Ferrara and A. Strominger in {\it Strings 89, Proceedings of
the Superstring Workshop, Texas A\&M University}, edited by
R. Arnowitt {\it et al} (World Scientific, 1989); \hfil\break
S. Cecotti, Comm. Math. Phys. 124 (1989) 23;
Comm. Math. Phys. 131 (1990) 517; \hfil\break
L.J. Dixon, V.S. Kaplunovsky and J. Louis,
\npb 329 (1990) 27; \hfil\break
A. Strominger, Comm. Math. Phys. 133 (1990) 163; \hfil\break
P. Candelas and X. De La Ossa, \npb 355 (1991) 455. \par
\refis{eff}
S. Ferrara, D. L\"ust, A. Shapere and S. Theisen,
\plb 225 (1989) 363; \hfil\break
S. Ferrara, D. L\"ust and S. Theisen,
\plb 233 (1989) 147; \hfil\break
A. Font, L.E. Ib\'a\~nez, D. L\"ust and F. Quevedo,
\plb 245 (1990) 401; \hfil\break
S. Ferrara, N. Magnoli, T.R. Taylor and G. Veneziano,
\plb 245 (1990) 409; \hfil\break
P. Bin\'etruy and M.K. Gaillard, \plb 253 (1991) 119; \hfil\break
H.P. Nilles and M. Olechowski, \plb 248 (1990) 268; \hfil\break
M. Cveti\v c, A. Font, L.E. Ib\'a\~nez, D. L\"ust and F. Quevedo,
\npb 361 (1991) 194; \hfil\break
L.E. Ib\'a\~nez and D. L\"ust, \plb 267 (1991) 51.\par
\refis{FKLZ}  S. Ferrara, C. Kounnas, D. L\"ust and F. Zwirner,
\npb 365 (1991) 431. \par
\refis{espejo} For recent reviews and references see \hfil\break
S. Ferrara, Mod. Phys. Lett. A6 (1991) 2175; \hfil\break
B.R. Greene and M.R. Plesser, ``Mirror Manifolds: A Brief
Review and Progress Report", preprint CLNS 91-1109 (1991);
``An Introduction to Mirror
Manifolds", preprint CLNS 91-1128 (1991).\par
\refis{c3}
Z. Maassarani, \plb 273 (1991) 457; \hfil\break
A. Klemm, S. Theisen and M. Schmidt, ``Correlation Functions for
Topological Landau-Ginzburg Models with $c \leq 3$", preprint
TUM-TP-129/91 (1991).\par
\refis{Landau} C. Vafa and N.P. Warner, \plb 218 (1989) 51;
\hfil\break
E. Martinec, \plb 217 (1989) 431.\par
\refis{CD} P. Candelas and X. De La Ossa,
\npb 355 (1991) 455. \par
\refis{SW} A. Strominger and E. Witten,
Comm. Math. Phys. 101 (1985) 341. \par
\refis{Stro} A. Strominger, Comm. Math. Phys. 133 (1990) 163. \par
\refis{BV} B. Blok and A. Varchenko, ``Topological Conformal Field
Theories and the Flat Coordinates", preprint
IASSNS-HEP-91/5 (1991). \par
\refis{CF} A.C. Cadavid and S. Ferrara, \plb 267 (1991) 193. \par
\refis{AL} P.S. Aspinwall and C.A. L\"utken, \npb 353 (1991) 427;
\npb 355 (1991) 482. \par
\refis{VW} E. Verlinde and N.P. Warner,  \plb 269 (1991) 96 .\par
\refis{LSW} W. Lerche, D.J. Smit and N.P. Warner, ``Differential
Equations for Periods and Flat Coordinates in Two Dimensional
Topological Matter Theories", preprint CALT-68-1738 (1991). \par
\refis{AM} P.S. Aspinwall and D.R. Morrison,
``Topological Field Theory and Rational Curves",
preprint DUK-M-91-12 (1991).\par
\refis{Morri1} D.R. Morrison,
``Mirror Symmetry and Rational Curves
in Quintic Threefolds: A Guide for Mathematicians",
preprint DUK-M-91-01 (1991). \par
\refis{Morri} D.R. Morrison,
``Picard-Fuchs Equations and Mirror
Maps for Hypersurfaces", preprint DUK-M-91-14 (1991). \par
\refis{FIQS} A. Font, L.E. Ib\'a\~nez, F. Quevedo and A. Sierra,
 \npb 337 (1989) 119. \par
\refis{Dwo} B. Dwork, Ann. of Math. 80 (1964) 227. \par
\refis{Gri1} R. Bryant and P.A. Griffiths, in {\it Arithmetic
and Geometry II}, edited by M. Artin and J. Tate
(Birkhauser, 1983). \par
\refis{Gri2} P.A. Griffiths, Ann. of Math. 90 (1969) 460. \par
\refis{Slater} L.J. Slater, {\it Generalized Hypergeometric Functions}
(Cambridge Univ. Press, 1966).\par
\refis{Ford} See for instance
L.R. Ford, {\it Automorphic Functions}, (Chelsea Publishing Co.,
1951).\par
\refis{Ince} See for instance E.L. Ince,
{\it Ordinary Differential Equations} (Dover, 1956). \par
\refis{Gepner87} D. Gepner, \npb 296 (1988) 757;
\plb 199 (1987) 380.\par
\refis{Lutken} C.A. L\"utken and G.G. Ross,
 \plb 213 (1988) 152; \hfil\break
M. Lynker and R. Schimmrigk, \plb 215 (1988) 681. \par
\refis{m0}
B. Greene, C. Vafa and N.P. Warner, \npb 324 (1989) 371;
\hfil\break
M. Lynker and R. Schimmrigk, \npb 339 (1990) 121. \par
\refis{Klein} F. Klein and R. Fricke, {\it Vorlesungen \"uber die
Theorie der elliptischen Modulfunktionen} (B.G. Teubner, 1890). \par
\endreferences

\vfill\eject

\bigskip\bigskip
\begintable
Model | $C_0$ | $C_1$ | $C_2$ | $C_3$ \cr
\ | \ | \ | \ | \ \crnorule
$A$ | $\ds{ {4\psi^2}\over {4\psi^6 - 1} }$
| $\ds{ {60\psi^3}\over {4\psi^6 - 1} }$
| $\ds{ {2(50\psi^6-1)}\over {\psi^2(4\psi^6 - 1)} }$
| $\ds{ {2(20\psi^6+1)}\over {\psi(4\psi^6 - 1)} }$ \crnorule
\ | \ | \ | \ | \ \cr
\ | \ | \ | \ | \ \crnorule
$B$ | $\ds{ 1\over {16(\psi^4 - 1)} }$
| $\ds{ {5\psi}\over {\psi^4- 1} }$
| $\ds{ {29\psi^2}\over {2(\psi^4- 1)} }$
| $\ds{ {8\psi^3}\over {\psi^4 - 1} }$ \crnorule
\ | \ | \ | \ | \ \cr
\ | \ | \ | \ | \ \crnorule
$C$ | $\ds{ {\psi}\over {4(4\psi^5 - 1)} }$
| $\ds{ {20\psi^2}\over {4\psi^5 - 1} }$
| $\ds{ {2(29\psi^5-1)}\over {\psi^2(4\psi^5 - 1)} }$
| $\ds{ {2(16\psi^5+1)}\over {\psi(4\psi^5 - 1)} }$  \crnorule
\ | \ | \ | \ | \
\endtable
\bigskip\bigskip
\leftline{{\bf Table 1.} \
Coefficients in the Picard-Fuchs Equation}

\bigskip\bigskip

\begintable
Model | $j$ | $\omega_j(\psi)$ \cr
 \ | \ | \ \crnorule
$A$ | 0,1,3,4 |
$ \psi^j F(\ds{{j+1}\over 6}, \ds{{j+1}\over 6},
\ds{{j+1}\over 6}, \ds{{j+1}\over 6};
\overbrace { {{j+2}\over {6}}, {{j+3}\over {6}},
{{j+5}\over {6}}, {{j+6}\over {6}} } ; 4\psi^6)$ \crnorule
 \ | \ | \ \cr
 \ | \ | \ \crnorule
$B$ | 0,1,2,3 |
$ \psi^j F(\ds{{2j+1}\over {8}}, \ds{{2j+1}\over {8}},
\ds{{2j+1}\over {8}}, \ds{{2j+1}\over {8}};
\overbrace { {{j+1}\over {4}}, {{j+2}\over {4}},
{{j+3}\over {4}}, {{j+4}\over {4}} } ; \psi^4)$ \crnorule
 \ | \ | \ \cr
 \ | \ | \ \crnorule
$C$ | 0,1,3,4 |
$\psi^j F(\ds{{2j+1}\over {10}}, \ds{{2j+1}\over {10}},
\ds{{2j+1}\over {10}}, \ds{{2j+1}\over {10}};
\overbrace { {{j+1}\over {5}}, {{j+2}\over {5}},
{{j+4}\over {5}}, {{j+5}\over {5}} } ; 4\psi^5)$ \crnorule
 \ | \ | \
\endtable

\bigskip\bigskip
\leftline{{\bf Table 2.} \
Solutions around $\psi = 0$}

\vfill\eject

\bigskip\bigskip

\begintable
 \ | \ \nr
$A$ | $\ds {y_0 = \psi^{-1} F({1\over 6}, {1\over 3},
{2\over 3}, {5\over 6}; 1,1,1; {1\over {4\psi^6}}) }$ \nr
 \ | \ \nr
 \ | $\ds{ y_1 = -y_0 \ln (6\psi) +
{{\psi^{-1}}\over 3} \sum_{l=0}^{\infty} \
{ {(6l)!}\over {(2l)! (l!)^4 (6\psi)^{6l} } } \left[
3\Psi(6l+1) - \Psi(2l+1) - 2\Psi(l+1) \right ] }$ \nr
\ | \ \cr
\ | \ \nr
$B$  | $\ds{ y_0 = \psi^{-{1\over 2}} F({1\over 8}, {3\over 8},
{5\over 8}, {7\over 8}; 1,1,1; {1\over {\psi^4}}) }$ \nr
\ | \ \nr
 \ | $\ds{ y_1 = -y_0 \ln (16\psi)
 + \psi^{-{1\over 2}}\sum_{l=0}^{\infty} \
{ {(8l)!}\over {(4l)! (l!)^4 (16\psi)^{4l} } } \left[
2\Psi(8l+1) - \Psi(4l+1) - \Psi(l+1) \right ] }$ \nr
\ | \ \cr
\ | \ \nr
$C$  | $\ds{ y_0= \psi^{-{1\over 2}} F({1\over {10}}, {3\over {10}},
{7\over {10}}, {9\over {10}}; 1,1,1; {1\over {4\psi^5}}) }$ \nr
\ | \ \nr
\ | $\ds{ y_1 = -y_0 \ln (20\psi) +
 {{\psi^{-{1\over 2}}}\over 5} \sum_{l=0}^{\infty} \
{ {(10l)!}\over {(5l)! (2l)! (l!)^3 (20\psi)^{5l} } } \left[
10\Psi(10l+1) - 5\Psi(5l+1) - 2\Psi(2l+1) - 3\Psi(l+1) \right ] }$ \nr
\ | \
\endtable
\bigskip\bigskip

\leftline{{\bf Table 3.} \
Solutions around $\psi = \infty$}

\bigskip\bigskip
\bigskip\bigskip\bigskip
\begintable
Model | $m_0$ | $m_1$ | $m_2$ | $m_3$ | $m_4$  \cr
$A$ | 3  |  7884  |  6028452  |
11900417220  |  34600752005688  \crnorule
$B$ | 2  |  29504  |  128834912  |
1423720546880 |  23193056024793312  \crnorule
$C$ | 1 | 231200 |  12215785600 |
1700894366474400 |  350154658851324656000
\endtable

\bigskip
\bigskip
\leftline{{\bf Table\  4.}
Coefficients in the expansion of the Yukawa couplings.}

\vfill\eject\bye